\documentclass[pra, amsmath,amssymb,nofootinbib,superscriptaddress,twocolumn]{revtex4-1}
\usepackage{bm,psfrag,graphicx,setspace,romannum}
\usepackage[T1]{fontenc}
\usepackage[left]{lineno}
\usepackage{xcolor}
\usepackage{hyperref}
\usepackage{bbold}
\usepackage{mathtools}
\usepackage{braket}
\usepackage{lineno}
\usepackage{soul}
\selectfont

\makeatletter
\def\l@subsubsection#1#2{}
\makeatother

\definecolor{myblue}{HTML}{4252cf}

\hypersetup{colorlinks=true, allcolors=myblue}

\newcommand{\RN}[1]{(\textrm{\Romannum{#1}})}

\begin{document}
\pagenumbering{arabic}
\title{Strongly coupled photonic molecules as doubly-coupled oscillators}
\author{Kevin C. Smith}
\altaffiliation{Present address: IBM Quantum}
\email{kcsmith@ibm.com}
\affiliation{Department of Physics, University of Washington, Seattle, Washington 98195, USA}
\author{Austin G. Nixon}
\affiliation{Department of Chemistry, University of Washington, Seattle, Washington 98195, USA}
\author{David J. Masiello}
\email{masiello@uw.edu}
\affiliation{Department of Chemistry, University of Washington, Seattle, Washington 98195, USA}

\begin{abstract}
In this work, we present a field-theoretic model of strongly coupled photonic molecules composed of interacting dielectric cavities in a closed, perfect-electric-conductor domain. Within this setting, we treat the resulting inter-mode couplings non-perturbatively. We demonstrate the predictive power of this framework by showing that supermode eigenfrequencies, field profiles, and mode volumes can be obtained directly from the isolated-cavity modes and dielectric environment, without electromagnetic simulations of the composite structure or numerical fitting. While our model affirms the phenomenological approach of modeling coupled cavity modes as simple coordinate-coupled oscillators in the weak coupling regime, we show that this intuition remarkably breaks down for strong coupling. Instead, we demonstrate that strongly coupled cavity modes are analogous to harmonic oscillators we term as \emph{doubly} coupled, with interactions via electric and magnetic fields appearing as independent coordinate-coordinate and momentum-momentum couplings, respectively. We show that this distinction is not merely cosmetic, but gives rise to observable properties while providing deep insights into the physical mechanism behind previously observed phenomena, such as coupling induced frequency shifts. Finally, we illustrate that the complex interplay of these dual couplings suggests the possibility to realize exotic phenomena that typically only occur in the ultrastrong coupling regime, here predicted to emerge for comparably modest mode splittings within a regime we term pseudo-ultrastrong coupling.
\end{abstract}
\maketitle


\section{Introduction}\label{sec:intro}
It is well known that the modes of an ideal electromagnetic cavity are the independent solutions to the homogeneous wave equation. While this statement follows trivially from Maxwell's equations, it has far reaching consequences which greatly simplify the study of systems involving optical cavities. In particular, it ensures that the cavity modes may be described in a separable fashion: i.e., the temporal and spatial dependences of the mode may be decoupled, ultimately leading to a time-dependent amplitude which obeys an equation of motion identical to that of a mass on a spring. In other words, the description of optical cavity modes may be reduced to a simple mechanical model of a harmonic oscillator. This not only greatly simplifies the study of systems with classical electromagnetic degrees of freedom, but also provides a clear path to quantization, famously exploited by Enrico Fermi in his widely adopted strategy for quantizing the radiation field \cite{Fermi1932}.

Among the innumerable applications of optical cavities, many have explicitly relied upon interactions between the photonic modes of adjacent cavities. A particularly influential example is the proposal by Yariv et al. to form coupled-resonator optical waveguides in order to achieve slowly propagating light for enhancement of nonlinear phenomena \cite{Yariv1999}. In the two decades following this formative work, numerous theoretical and experimental investigations have explored applications of so-called photonic molecules -- systems composed of a finite number of coupled dielectric cavities, named in analogy to their atomic counterparts. The applications of photonic molecules are wide-ranging~\cite{Liao_2020} and include, for example, low-threshold lasing \cite{Nakagawa2005, Boriskina2006a}, electromagnetic-induced transparency \cite{Xu2006, Smith2004, Yang2009}, nonclassical light generation \cite{Liew2010, Bamba2011, Dousse2010, Saxena2019}, quantum simulation \cite{Underwood2012, Majumdar2012, Hartmann2016}, and parity-time symmetry \cite{Peng2014, Chang2014a}.

Because individual dielectric cavity modes are often modeled through their isomorphism to harmonic oscillators, it then stands to reason that systems of electromagnetically interacting cavities must be well described by coupled oscillator equations. Such is the idea of time-dependent coupled mode theory (CMT) \cite{Haus1991,haus1984waves}, a heuristic workhorse which has been used near-ubiquitously in experimental and theoretical investigations of photonic molecules. While it has proved to be an invaluable tool for simple modeling of generic coupled cavity systems, CMT is phenomenological in nature, often relying on numerical fits to either simulation or experimental data to determine model parameters. Furthermore, CMT is an approximate description of the underlying physics, agreeing with first-principles electromagnetic theory only in the weak\footnote{Here, weak coupling means that the coupling strength is small relative to the resonance frequency of the cavity mode considered. This distinction is important, as the term weak coupling in the cavity QED literature refers to a comparison between the coupling strength and the largest rate of dissipation in the system.} coupling regime \cite{Haus1991}, thus limiting its usefulness for strongly coupled photonic molecules. As a result, CMT does not by itself provide the freedom of analytic exploration and predictivity desired for modern applications which depend upon an understanding of strongly coupled cavity phenomena at a high level. Thus, a first-principles theory of photonic molecules which (i) is applicable beyond the weak coupling approximation and (ii) can be used to theoretically predict supermode properties without numerical fitting is highly desirable.

Toward this goal, here we present a field-theoretic framework for describing strongly coupled photonic molecules consisting of two or more dielectric cavities, each supporting a finite set of spectrally isolated modes with discrete labels. For analytic simplicity, we model these as lossless modes within a large, finite domain subject to perfect-electric-conductor (PEC) boundary conditions at the domain boundaries, i.e., the infinite-$Q$ idealization. For the near-field observables of interest in this work, this construction serves as a convenient proxy for spectrally isolated, high-$Q$ dielectric modes, whose leakage can then be incorporated perturbatively. We approach this problem using techniques of Lagrangian and Hamiltonian mechanics, a framework which is readily adaptable to modern applications in quantum science and engineering due to ease of quantization and facile extension to include nonlinear quantum emitters such as defect centers or and quantum dots \cite{Majumdar2012, Saxena2019, Liao_2020}. Notably, we treat the inter-cavity coupling non-perturbatively, going beyond the weak-coupling approximations of CMT. We illustrate the predictive power of this theory, demonstrating the ability to compute properties of the supermodes of the composite structure given knowledge only of its constituent components, circumventing the need for costly electromagnetic simulations of the full structure. Consequently, this theory provides a route to scalably explore supermode properties in complex photonic molecules through analytical and numerical means.

In addition to its practical utility, we show that this field-theoretic model reveals an unexpected yet fascinating insight into the physics underlying interacting cavity modes: it suggests that the very intuition CMT relies on -- that coupled cavity modes are akin to coordinate-coupled oscillators -- is only approximate. Rather, we show that coupled cavity modes behave as oscillators which are ``doubly'' coupled – i.e., both through their coordinates and velocities (or momenta) independently, leading to notable classical and quantum mechanical deviations from the “singly” coupled oscillators that our intuition is built upon. In addition to identifying the limiting case in which this generalized coupling is reducible to a single parameter, we show that its full consideration explains, from first principles, previously identified effects such as coupling-induced resonance frequency shifts, which are not evident in CMT without introducing phenomenological self-coupling parameters \cite{Popovic2006}. Finally, we conclude by demonstrating that the interplay of these dual couplings suggests the possibility for exotic phenomena typically realizable only in the experimentally challenging ultrastrong coupling regime, here predicted to be accessible in photonic molecules within a comparatively modest parameter regime which we term \textit{pseudo-ultrastrong coupling} (pUSC).

The subsequent sections are organized as follows: In Section \ref{sec:lagrangian} we develop a field-theoretic formalism for coupled dielectric cavities in a PEC-enclosed domain, presented through the lens of Lagrangian mechanics. In Section \ref{sec:singlemode}, we reduce this general theory to the case of two single-mode cavities and demonstrate that a careful treatment of the electromagnetic interactions in such a system leads to coupled mode equations which we term \emph{doubly-coupled oscillators} (DCOs). In Section \ref{subsec:supermode} we derive closed expressions for supermode properties of interest such as mode functions, mode volumes, and and resonance frequencies, illustrating the predictive power of this formalism and its potential benefit over expensive electromagnetic simulations of the composite structure. In Section \ref{subsec:weakcoupling}, we show how our DCO model reduces to the more typical \emph{coordinate-coupled oscillators} (CCOs) in the limit of weak coupling, consistent with the intuition of CMT. Following this, in Section \ref{subsec:PUSC} we show that quantization of the DCO model suggests the possibility to experimentally realize exotic phenomena typically only associated with ultrastrong coupling. Finally, in Section \ref{subsec:nanobeams} we provide a simple example of our theory applied to a system of two nanobeam resonators. Section \ref{sec:conclusion} summarizes our findings.

\section{The coupled cavity Lagrangian}\label{sec:lagrangian}
\subsection{Single dielectric cavity}\label{ssec:single_cavity}
We begin by considering a single dielectric cavity and show how it may be mapped onto a Lagrangian corresponding to a set of independent harmonic oscillators. The electric and magnetic fields of a dielectric cavity obey the macroscopic sourceless Maxwell's equations,
\begin{equation}
\begin{split}
&\nabla\cdot\varepsilon(\mathbf{r})\mathbf{E}=0 \\
&\nabla\cdot\mathbf{B}=0 \\
&\nabla\times\mathbf{E}=-\frac{1}{c}\dot{\mathbf{B}} \\
&\nabla\times\mathbf{B}=\frac{1}{c}\varepsilon(\mathbf{r})\dot{\mathbf{E}},
\end{split}
\label{eq:macromax}
\end{equation}
where the permeability $\mu=1$ has been enforced as only non-magnetic materials are of interest in this work, and the spatial and temporal dependence of $\mathbf{E}$ and $\mathbf{B}$ is implied. The inhomogeneous dielectric function $\varepsilon(\mathbf{r})$, here assumed to be real valued and dispersionless, accounts for contributions to the fields due to the polarizable media which support the cavity. 

While Maxwell's equations provide sufficient information for solving the modes of a particular cavity, it is convenient to cast them in terms of the potentials defined by the usual relations $\mathbf{E} = -\dot{\mathbf{A}}/c-\nabla\phi$ and $\mathbf{B} = \nabla\times\mathbf{A}$. The most obvious advantage of this reformulation is that two of Maxwell's equations are automatically satisfied by these relations, therefore reducing the total number of coupled partial differential equations which must be simultaneously solved to two. Furthermore, the relevant degrees of freedom can then be identified as the potentials and their time derivatives, leading to a description of the system via generalized coordinates and velocities amenable to Lagrangian and Hamiltonian formalisms and, consequently, canonical quantization. This reformulation comes at a cost, however, as redundancies arise in the description and must be properly removed. 

In nonrelativistic quantum electrodynamics, this is typically achieved through (i) specialization to the Coulomb gauge ($\nabla\cdot\mathbf{A}=0$) and (ii) subsequent algebraic elimination of the scalar potential from the electromagnetic Lagrangian \cite{cohen1997photons}. The primary feature of the Coulomb gauge is the complete dependence of the scalar potential on matter degrees of freedom. Correspondingly, the vector potential alone encodes the true dynamical degrees of freedom of the electromagnetic field and, in the absence of matter, the scalar potential may be taken to zero without loss of generality. 

In the present case, however, we are not interested in the free-space electromagnetic degrees of freedom, but rather those supported by an electromagnetic cavity composed of bound matter characterized by the macroscopic dielectric function $\varepsilon(\mathbf{r})$. The analog to the Coulomb gauge in this context is the so-called generalized Coulomb gauge, defined by the condition $\nabla\cdot\varepsilon(\mathbf{r})\mathbf{A}=0$ \cite{Glauber1991,Dalton1996}. With this choice, the scalar potential becomes entirely dependent upon the \emph{free} matter degrees of the system and may therefore be taken to zero without loss of generality. Consequently, Maxwell's equations reduce to just a single partial differential equation: the generalized wave equation for the vector potential
\begin{equation}
\nabla\times\nabla\times\mathbf{A} + \frac{\varepsilon(\mathbf{r})}{c^2}\ddot{\mathbf{A}}=0,
\label{eq:waveeq}
\end{equation}
the solutions of which fully encode the complete set of cavity modes in the dielectric environment described by $\varepsilon(\mathbf{r})$. The vector potential may therefore be written as a sum over these independent solutions as
\begin{equation}
\mathbf{A}(\mathbf{r},t) = \sum_m\frac{c\sqrt{4\pi}}{V_m}q_m(t)\mathbf{f}_m(\mathbf{r}),
\label{eq:expansion}
\end{equation}
where $q_m(t)$ is a time-dependent amplitude, $\mathbf{f}_m(\mathbf{r})$ is the mode function, and $V_m$ is the mode volume. Crucially, the mode functions are solutions to the generalized Helmholtz equation
\begin{equation}
	\nabla\times\nabla\times\mathbf{f}_m(\mathbf{r}) = \frac{\omega_m^2}{c^2}\varepsilon(\mathbf{r})\mathbf{f}_m(\mathbf{r}),
\label{eq:helmholtz}
\end{equation}
 where $\omega_m$ is the resonant frequency of the $m$th mode.

Many of the properties of the mode functions $\mathbf{f}_m(\mathbf{r})$ can be determined by introducing the rescaled functions $\mathbf{u}_m(\mathbf{r}) = \sqrt{\varepsilon(\mathbf{r})}\mathbf{f}_m(\mathbf{r})$ which are eigenfunctions of the symmetrized Helmholtz operator \cite{Glauber1991}:
\begin{equation}
\label{eq:doublecurlu}
\frac{1}{\sqrt{\varepsilon(\mathbf{r})}}\nabla\times\nabla\times\frac{1}{\sqrt{\varepsilon(\mathbf{r})}}\mathbf{u}_m(\mathbf{r})=\frac{\omega_m^2}{c^2}\mathbf{u}_m(\mathbf{r}).
\end{equation}

In the derivations that follow, we work on a large but finite domain $\mathcal{V}$ with PEC boundary conditions, for which the symmetrized Helmholtz operator is Hermitian. In this idealized setting, the mode functions $\mathbf{f}_m(\mathbf{r})$ are normalizable on $\mathcal{V}$ and can be chosen to be real without loss of generality, and the functions $\mathbf{u}_m(\mathbf{r})$ form an orthogonal set. This infinite-$Q$ construction serves as a proxy for the realistic high-$Q$, spectrally isolated dielectric modes of interest here, such as those realized in nanobeam photonic crystals and ring resonators~\cite{Majumdar2012,Smith2020}. In this setting, radiative loss can be incorporated perturbatively, as in conventional CMT \cite{haus1984waves}. While beyond the scope of the present work, we note that a full quasinormal-mode (QNM) treatment of open, radiating resonators \cite{Kristensen2014,Kristensen2015, ren2021quasinormal, Sauvan_2022, wu2024exact} could additionally enable rigorous predictions for radiative observables in cavities of arbitrary quality $Q$; here we focus on near-field quantities for which this PEC idealization is a good approximation, provided the modes are sufficiently high-$Q$.

Fixing the normalization of $\mathbf{f}_m(\mathbf{r})$ at the expense of rescaling the amplitude $q_m(t)$, the mode functions are then endowed with following set of properties:

\begin{subequations}
\label{eq:modeproperties}
\begin{enumerate}
\item \emph{Normalization:} $\mathbf{f}_m(\mathbf{r})$ is normalized such that $\textrm{max} \{\varepsilon(\mathbf{r})\left|\mathbf{f}_m\right|^2\}=1$, and therefore the mode volume is naturally defined by
\begin{equation}
V_m = \frac{\int_{\mathcal{V}} d^3r\,\varepsilon(\mathbf{r})\left|\mathbf{E}_m(\mathbf{r})\right|^2}{\textrm{max} \{\varepsilon(\mathbf{r})\left|\mathbf{E}_m(\mathbf{r})\right|^2\}} = \int_{\mathcal{V}} d^3r\,\varepsilon(\mathbf{r})\left|\mathbf{f}_m(\mathbf{r})\right|^2,
\label{eq:normalization}
\end{equation}
where $\mathbf{E}_m(\mathbf{r})$ is the electric field contributed by the $m$th mode.
\item \emph{Orthogonality:} The set of mode functions $\{\mathbf{f}_m(\mathbf{r})\}$ are orthogonal:
\begin{equation}
\int_{\mathcal{V}} d^3r\, \varepsilon(\mathbf{r})\mathbf{f}_m(\mathbf{r})\cdot\mathbf{f}_n(\mathbf{r})=V_m\delta_{mn}
\label{eq:orthogonality}
\end{equation}
\item \emph{Completeness on $\mathcal{V}$:} Within the PEC-enclosed domain $\mathcal{V}$, the mode functions $\mathbf{f}_m(\mathbf{r})$ furnish a resolution of the identity on the subspace of functions obeying the transversality condition,
\begin{equation}
\nabla\cdot\varepsilon(\mathbf{r})\mathbf{f}_m(\mathbf{r})=0,
\label{eq:transverse}
\end{equation}
i.e., on the $\varepsilon$-transverse subspace of $L^2(\mathcal V)$. Equivalently, one may introduce the generalized transverse $\delta$-function \cite{Glauber1991,cohen1997photons}
\begin{equation}
\delta^{\varepsilon\perp}_{ij}(\mathbf{r},\mathbf{r}') = \sum_m\frac{1}{V_m}(\mathbf{f}_{m}(\mathbf{r})\cdot\hat{\mathbf{e}}_i)(\mathbf{f}_{m}(\mathbf{r}')\cdot\hat{\mathbf{e}}_j),
\end{equation}
which, via the integral equation 
\begin{equation}
\mathbf{V}_i^{\varepsilon\perp}(\mathbf{r})=\int_{\mathcal{V}} d^3r'\,\varepsilon(\mathbf{r}')\delta_{ij}^{\varepsilon\perp}(\mathbf{r},\mathbf{r'})\mathbf{V}_j(\mathbf{r}'),
\end{equation}
projects any vector field $\mathbf V(\mathbf r)$ onto its $\varepsilon$-transverse component $\mathbf{V}^{\varepsilon\perp}(\mathbf{r})$ satisfying $\nabla\cdot\varepsilon(\mathbf{r})\mathbf{V}^{\varepsilon\perp}(\mathbf{r})=0$. We emphasize that this projection is a finite-domain device tied to the PEC idealization; in truly open systems with outgoing-wave boundary conditions the physical resonant (quasinormal) modes are not square-integrable under the standard inner product and therefore do not compose an $L^2$-complete basis \cite{Kristensen2014, Sauvan_2022}.

\end{enumerate}
\end{subequations}

The dynamical equations which govern the behavior of the independent cavity modes may be computed using various methods. One option is to express the wave equation in Eq.~\eqref{eq:waveeq} in terms of the expansion in Eq.~\eqref{eq:expansion}, multiply by $\mathbf{f}_n(\mathbf{r})$ and integrate, exploiting orthogonality to reveal independent equations of motion for each mode amplitude $q_n(t)$. Here we take an alternate approach by appealing to Lagrangian mechanics. While the same equations of motion result through either strategy, the latter is ultimately more flexible as it provides a route for computing the Hamiltonian and is therefore amenable to quantization. Furthermore, a Hamiltonian (or Lagrangian) based approach simplifies extension of the model to include additional components such as emitters, as well as dissipation through inclusion of system-bath interaction terms via standard methods.

In the absence of free charge, the electromagnetic Lagrangian in the generalized Coulomb gauge is given by \cite{cohen1997photons}
\begin{equation}
L = \int_{\mathcal{V}} \frac{d^3r}{8\pi}\left[\varepsilon(\mathbf{r})\frac{\dot{\mathbf{A}}^2}{c^2} - (\nabla\times\mathbf{A})^2\right],
\label{eq:EMlagrangian}
\end{equation}
where the integration volume corresponds to the domain $\mathcal{V}$ containing the cavity and its dielectric surroundings (see discussion below Eq.~\eqref{eq:doublecurlu}).  Expanding $\mathbf{A}$ according to Eq.~\eqref{eq:expansion} and integrating the second term by parts, it can be shown \cite{smith2021theoretical} 
 that the equation of motion for the amplitude of the $m$th cavity mode is
\begin{equation}
\frac{1}{V_m}\ddot{q}_m + \frac{\omega_m^2}{V_m}q_m=0,
\end{equation}
where all spatial dependence has been integrated out via application of the orthogonality relation Eq.~\eqref{eq:orthogonality}. This result suggests that the dynamics of a set of independent cavity modes are equivalent to that of a set of independent harmonic oscillators. While this correspondence is well known, we here reestablish this result to provide intuition and serve as a foundation for generalization to the more interesting coupled cavity case. One unique aspect of our formulation is the correspondence between the inverse of the mode volume $V_m$ and the oscillator effective mass. This connection is not typically recognized as most often the mode functions are normalized to unity rather than the normalization condition chosen in Eq.~\eqref{eq:normalization}. Regardless, this analogy will later be relied upon more explicitly in computing the mode volumes of photonic molecule supermodes.

\subsection{Gauge transformation of the isolated cavity modes}
\label{subsec:gaugetrans}
There are two independent strategies for extending the methods of the previous section to the case of coupled dielectric cavities. The first relies on the realization that the system, while consisting of multiple cavities which may be viewed individually, as a whole must still be described by some total dielectric function $\varepsilon(\mathbf{r})$. Consequently, the supermodes are determined by solving the generalized wave equation Eq.~\eqref{eq:waveeq} with the full dielectric function substituted for that of an individual cavity, and all of the resulting properties discussed in the preceding section correspondingly follow. This approach has obvious drawbacks, however, as $\varepsilon(\mathbf{r})$ may describe a set of independent cavities which together form a sufficiently large and complex system such that numerically solving Eq.~\eqref{eq:waveeq} is computationally challenging. Even in the case of a dimer, heterogeneity of the photonic molecule can inhibit simulation of the composite structure due to, for example, mismatched length scales of the composing cavities \cite{thakkar2017sculpting, pan2019elucidating, Smith2020}. Furthermore, changing the gap size between adjacent cavities redefines $\varepsilon(\mathbf{r})$ and therefore alters Eq.~\eqref{eq:waveeq}, necessitating a whole new set of numerical calculations. As a result, exploration of the influence of inter-cavity separation and orientation on the properties of supermodes becomes extremely costly, if not outright prohibitive, through this strategy.

A second more flexible approach involves solving for the modes of the individual cavities and analytically blending them to form supermodes. In contrast to the previously described route, this strategy is efficiently scalable as Eq.~\eqref{eq:waveeq} only needs to be solved for a single cavity at a time, reducing the computational complexity of solving for the supermodes to the diagonalization of $N\times N$ matrices, where $N$ is the number of independent cavity modes under consideration. Furthermore, it allows for the numerical and (depending upon the size and symmetries of the system) analytic computation of supermode resonant frequencies, mode functions, and mode volumes. This in turn provides a route to understand and predict the dependence of supermode properties on geometric parameters such as the relative position and orientation between adjacent cavities, all without requiring expensive electromagnetic simulations on the full system. 

Before further developing our divide-and-conquer approach, we first address a subtlety which underpins the theory presented here, relating to relative change of dielectric environment between the single and multiple cavity case. Recalling the single cavity formalism of the previous section, both the vector potential and the electric field obey identical transversality conditions, the former via the generalized Coulomb gauge ($\nabla\cdot\varepsilon(\mathbf{r})\mathbf{A}=0$), and the latter due to Gauss's law ($\nabla\cdot\varepsilon(\mathbf{r})\mathbf{E}=0$). As a result, both $\mathbf{A}$ and $\mathbf{E}$ may be expanded in terms of the same set of basis functions $\mathbf{f}_m(\mathbf{r})$ which solve Eq.~\eqref{eq:helmholtz}. This is unsurprising as the vector potential and electric field are simply related by $\mathbf{E}=-\dot{\mathbf{A}}/c$ in the absence of free charge. 

Nuances arise, however, when a second dielectric cavity is added to the system. Momentarily ignoring the additional electromagnetic degrees of freedom of the second cavity, the dielectric environment containing both cavities is inevitably different from that of the single cavity by some function $\delta\varepsilon(\mathbf{r})$. Unavoidably, this leads to an altered transversality condition on $\mathbf{E}$ and the functions $\mathbf{f}_m(\mathbf{r})$ therefore no longer form an appropriate basis for expansion of the electric field of the first cavity, a consequence which was previously explored in the context of perturbation theory of Maxwell's equations in Ref.~\cite{Johnson2002}. A physically intuitive explanation for this failure is that the change in dielectric media contributed by the second cavity $\delta\varepsilon(\mathbf{r})$, once polarized by the field of the first cavity, acts like ``free charge'' in the gauge $\nabla\cdot\varepsilon(\mathbf{r})\mathbf{A}=0$ and therefore contributes to the scalar potential. In other words, the electric field and vector potential are no longer simply related by a time derivative and, instead, the electric field contains additional longitudinal contributions (i.e., $\mathbf{E} = -\dot{\mathbf{A}}/c - \nabla\phi$) that the unperturbed mode functions $\mathbf{f}_m(\mathbf{r})$ cannot describe alone. 

Fortunately, this complication may be preempted through gauge transformation of the single cavity mode expansion. As summarized in Figure \ref{fig:f1} and discussed in greater detail in Appendix \ref{app:gauge}, there is a subtle connection between the choice of gauge, the representation of the dielectric environment, and the appearance of free charge. When writing the macroscopic form of Maxwell's equations, one must choose whether to represent matter as either polarizable media described by some dielectric function, or as free matter. Oftentimes, the most appropriate solution is to partition the matter such that the system contains both, as is typical in formulations of quantum electrodynamics in dielectric media \cite{Dalton1996, Hillery1984}. Regardless of how the system is partitioned, there always exists a corresponding generalized Coulomb gauge, itself dependent on the choice of dielectric function such that the scalar potential is reduced to complete dependency upon the remaining free charge. This statement alone illustrates how the representation of dielectric environment, free charge, and gauge are all intertwined, and it is therefore unsurprising that one may effectively ``repartition'' the system by transforming between gauges of the form $\nabla\cdot\xi(\mathbf{r})\mathbf{A}=0$, as summarized in Figure \ref{fig:f1} and shown explicitly in Appendix \ref{app:gauge}. With this in mind, the manifestation of a nonzero scalar potential upon modification of the single cavity dielectric function may be preempted through generalization of the mode expansion in Eq.~\eqref{eq:expansion} to the most general Coulomb-like gauge $\nabla\cdot\xi(\mathbf{r})\mathbf{A}=0$, where $\xi(\mathbf{r})$ is a placeholder for an arbitrary dielectric function. 

\begin{figure}
\centering
\includegraphics[scale = 0.7]{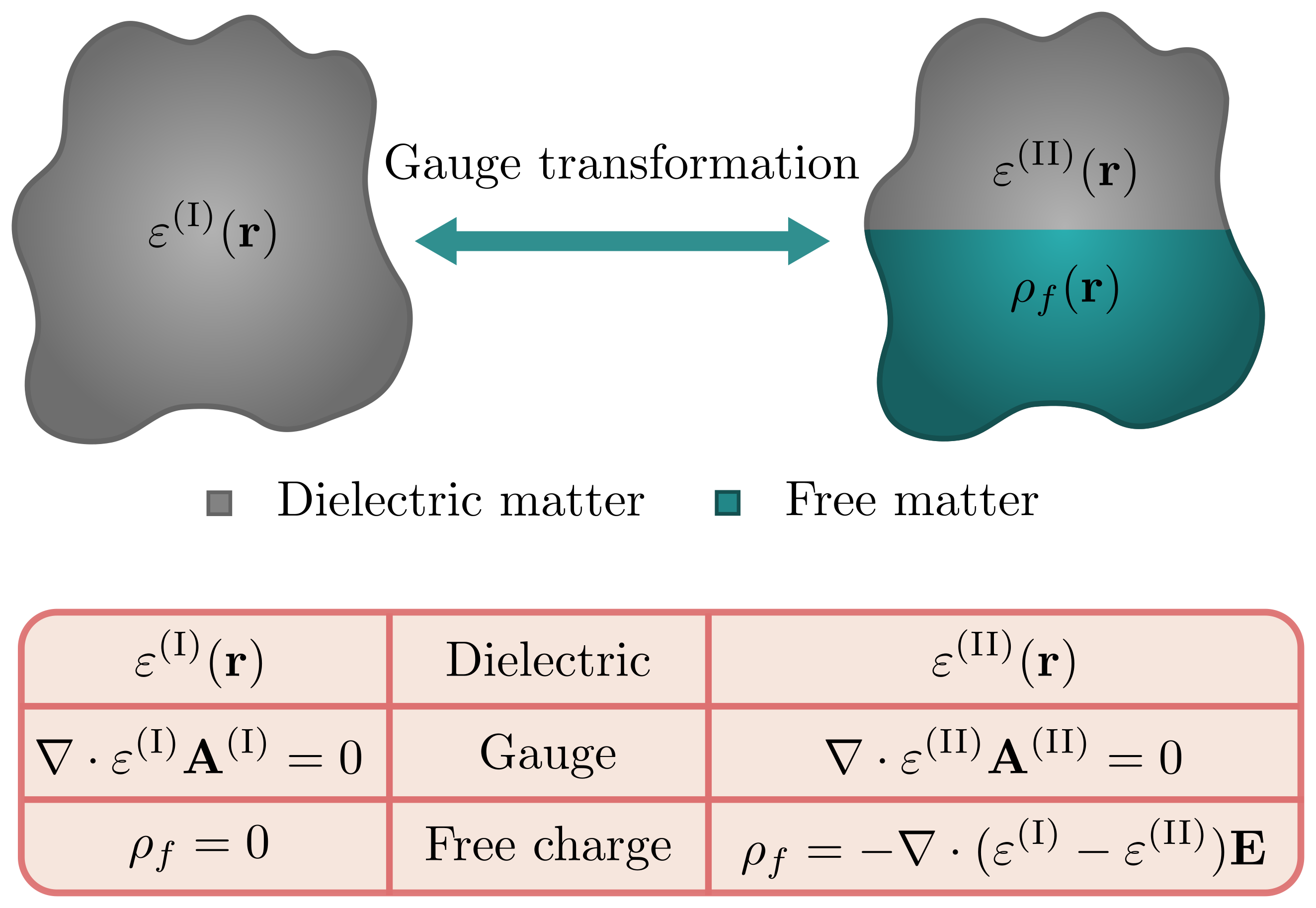}
\caption{Alternate representations of the dielectric environment through gauge transformation. Left and right illustrations display an identical cluster of matter partitioned in two distinct ways, labelled $\RN{1}$ and $\RN{2}$. In formulation $\RN{1}$, all of the matter is packaged into the dielectric function $\varepsilon^{\RN{1}}(\mathbf{r})$ with none represented as free charge. In formulation $\RN{2}$, the matter is partitioned into the dielectric function $\varepsilon^{\RN{2}}(\mathbf{r})$ and the free charge $\rho_f=-\nabla\cdot(\varepsilon^{\RN{1}}-\varepsilon^{\RN{2}})\mathbf{E}$. As shown in Appendix \ref{app:gauge}, the potentials in each formulation are equal up to gauge transformation, revealing the intimate link between representation of the dielectric environment, free charge, and Coulomb-like gauges defined by $\nabla\cdot\xi(\mathbf{r})\mathbf{A}=0$.}
\label{fig:f1}
\end{figure}

Returning to the single cavity expansion of Eq.~\eqref{eq:expansion} and recalling that $\phi=0$ in the single cavity dielectric environment $\varepsilon(\mathbf{r})$, gauge transformation results in the new potentials $\mathbf{A}' = \mathbf{A} + \nabla \Lambda$ and $\phi' = -\dot{\Lambda}/c$, where $\Lambda(\mathbf{r},t)$ is an arbitrary time-dependent scalar field. Establishing the new gauge condition $\nabla\cdot\xi(\mathbf{r})\mathbf{A}'=0$ constrains $\Lambda$ to the set of functions which obey $\nabla\cdot\xi(\mathbf{r})\nabla \Lambda=\nabla\cdot[\varepsilon(\mathbf{r})-\xi(\mathbf{r})]\mathbf{A}$. While this constraint appears somewhat abstract at first glance, taking a time derivative of both sides leads to the generalized Poisson equation
\begin{equation}
\nabla\cdot\xi(\mathbf{r})\nabla\phi'=\nabla\cdot[\varepsilon(\mathbf{r})-\xi(\mathbf{r})]\mathbf{E},
\end{equation}
clearly revealing $\xi(\mathbf{r})$ as a new, ``effective'' dielectric environment. Likewise, the right hand side appears as a source term resulting from the polarization of media in the region where $\varepsilon(\mathbf{r})-\xi(\mathbf{r})\neq 0$, and is equivalent to the effective free charge $\rho_f=-\nabla\cdot\mathbf{P}$ where $\mathbf{P}=[\varepsilon(\mathbf{r})-\xi(\mathbf{r})]\mathbf{E}/4\pi$.

In order to simplify the mode expansion for $\mathbf{A}'$ as much as possible, it is convenient to expand $\Lambda$ as
\begin{equation}
\Lambda(\mathbf{r},t)=\sum_m\frac{c\sqrt{4\pi}}{V_m}q_m(t)\psi_m(\mathbf{r}),
\end{equation}
where $q_m(t)$ is the mode amplitude appearing in Eq.~\eqref{eq:expansion}. Correspondingly, the gauge transformed vector and scalar potentials can be written as
\begin{equation}
\begin{split}
\mathbf{A}'(\mathbf{r},t) &= \sum_m\frac{c\sqrt{4\pi}}{V_m}q_m(t)\widetilde{\mathbf{f}}_m(\mathbf{r}) \\
\phi'(\mathbf{r},t) &= -\sum_m\frac{\sqrt{4\pi}}{V_m}\dot{q}_m(t)\psi_m(\mathbf{r}),
\end{split}
\label{eq:newpot}
\end{equation}
where $\widetilde{\mathbf{f}}_m(\mathbf{r}) = \mathbf{f}_m(\mathbf{r}) + \nabla\psi_m(\mathbf{r})$ is the gauge generalized mode function and $\nabla\cdot\xi(\mathbf{r})\nabla\psi_m(\mathbf{r})=\nabla\cdot[\varepsilon(\mathbf{r})-\xi(\mathbf{r})]\mathbf{f}_m(\mathbf{r})$. While this gauge transformation leaves the fields unaltered and therefore changes very little in terms of the single cavity, the resulting Coulomb-like gauge condition $\nabla\cdot\xi(\mathbf{r})\mathbf{A}'=0$ is more amenable toward extension to systems of coupled dielectric cavities which share a dielectric environment distinct from that of the individual cavities in isolation, as will be further demonstrated in the next section.

\subsection{Two coupled dielectric cavities}
With the gauge generalized description of a single cavity in hand, extension of the theory to coupled dielectric cavities is straightforward. For simplicity, we apply this formalism for the situation of a cavity dimer, but emphasize that the outlined procedure generalizes for $N$ cavities. As a starting point, we assume that each cavity has a well defined single cavity dielectric function $\varepsilon_i(\mathbf{r})$, where $i=1,2$ labels each cavity. For the remainder of this work, $\varepsilon_i(\mathbf{r})$ will indicate the dielectric function of the $i$th cavity isolated, while $\varepsilon(\mathbf{r})$ represents the full dielectric function of the composite dimer.

Similar to the single cavity case, the scalar potential may be eliminated through specialization to the generalized Coulomb gauge $\nabla\cdot\varepsilon(\mathbf{r})\mathbf{A}=0$, ensuring that both $\mathbf{E}$ and $\mathbf{A}$ may be expanded in the same basis in the absence of free charge. Relying on the discussion of the previous section, the gauge generalized single cavity mode functions introduced in Eq.~\eqref{eq:newpot} provide such a basis upon the substitution $\xi(\mathbf{r})\rightarrow\varepsilon(\mathbf{r})$. Accordingly, the total vector potential may be written as 
\begin{equation}
\mathbf{A}(\mathbf{r},t) = \sum_{im}\frac{c\sqrt{4\pi}}{V_{im}}q_{im}(t)\widetilde{\mathbf{f}}_{im}(\mathbf{r}),
\label{eq:twocavityexpansion}
\end{equation}
where $i=1,2$ denotes each cavity of the dimer and $\widetilde{\mathbf{f}}_{im}(\mathbf{r}) = \mathbf{f}_{im}(\mathbf{r})+ \nabla\psi_{im}(\mathbf{r})$ is the gauge generalized $m$th mode function of the $i$th cavity. Its contributions consist of both the ``bare'' mode function $\mathbf{f}_{im}(\mathbf{r})$ of the (isolated) $i$th cavity and the longitudinal correction $\nabla\psi_{im}(\mathbf{r})$, describing the field contributed via polarization of the ``added'' media $ \varepsilon(\mathbf{r}) - \varepsilon_i(\mathbf{r})$ in accordance with the generalized Poisson equation,
\begin{equation}
\nabla\cdot\varepsilon(\mathbf{r})\nabla\psi_{im}(\mathbf{r})=-\nabla\cdot[\varepsilon(\mathbf{r})-\varepsilon_i(\mathbf{r})]\mathbf{f}_{im}(\mathbf{r}).
\label{eq:genpoisson}
\end{equation} 
Crucially, the bare mode functions $\mathbf{f}_{im}(\mathbf{r})$ may be found by solving the single cavity generalized Helmholtz equation Eq.~\eqref{eq:helmholtz}, while the longitudinal corrections can be computed by numerically solving Eq.~\eqref{eq:genpoisson}. The gauge generalized mode functions $\widetilde{\mathbf{f}}_{im}(\mathbf{r})$ are therefore entirely determinable from the modal decomposition and dielectric function of the \emph{isolated} constituent cavities, in combination with the dielectric function of the composite photonic molecule.

In analogy to the preceding single cavity analysis, we formulate equations of motion for the coupled cavity system using the electromagnetic Lagrangian in Eq.~\eqref{eq:EMlagrangian}, now reexpressed in terms of the dimer mode expansion of Eq.~\eqref{eq:twocavityexpansion}:
\begin{widetext}
\begin{equation}
\begin{split}
L = &\frac{1}{2}\sum_{imn}\frac{\dot{q}_{im}\dot{q}_{in}}{V_{im}V_{in}}\left[V_{im}\delta_{mn}+\Sigma_{imn}\sqrt{V_{im}V_{in}}\right] - \frac{1}{2}\sum_{im}\frac{\omega_{im}^2}{V_{im}}q_{im}^2 \\
&+\sum_{mn}\frac{1}{\sqrt{\omega_{1m}\omega_{2n}V_{1m}V_{2n}}}g_{mn}^{(E)}\dot{q}_{1m}\dot{q}_{2n} -\sum_{mn}\sqrt{\frac{\omega_{1m}\omega_{2n}}{V_{1m}V_{2n}}}g_{mn}^{(M)}q_{1m}q_{2n},
\end{split}
\label{eq:Lagrangian}
\end{equation}
where the intracavity couplings ($\Sigma_{imn}$) and intercavity electric ($g_{mn}^{(E)}$) and magnetic ($g_{mn}^{(M)}$) 
 couplings are defined by
\begin{equation}
\begin{split}
	\Sigma_{imn} &= \frac{1}{\sqrt{V_{im}V_{in}}}\int_{\mathcal{V}} d^3r [\varepsilon(\mathbf{r})-\varepsilon_i(\mathbf{r})]\mathbf{f}_{im}(\mathbf{r})\cdot\widetilde{\mathbf{f}}_{in}(\mathbf{r}) \\
	g_{mn}^{(E)} &= \sqrt{\frac{\omega_{1m}\omega_{2n}}{V_{1m}V_{2n}}}\int_{\mathcal{V}} d^3r \varepsilon(\mathbf{r})\left[\mathbf{f}_{1m}(\mathbf{r})\cdot\mathbf{f}_{2n}(\mathbf{r}) - \nabla\psi_{1m}(\mathbf{r})\cdot\nabla\psi_{2n}(\mathbf{r})\right] \\
	g_{mn}^{(M)} &= \frac{1}{2}\sqrt{\frac{1}{\omega_{1m}\omega_{2n}V_{1m}V_{2n}}}\int_{\mathcal{V}} d^3r \left[\omega_{1m}^2\varepsilon_1(\mathbf{r}) + \omega_{2n}^2\varepsilon_2(\mathbf{r})\right]\mathbf{f}_{1m}(\mathbf{r})\cdot\mathbf{f}_{2n}(\mathbf{r}).
\end{split}
\label{eq:couplings}
\end{equation}
\end{widetext}

\begin{figure}
\centering
\includegraphics[scale =0.8]{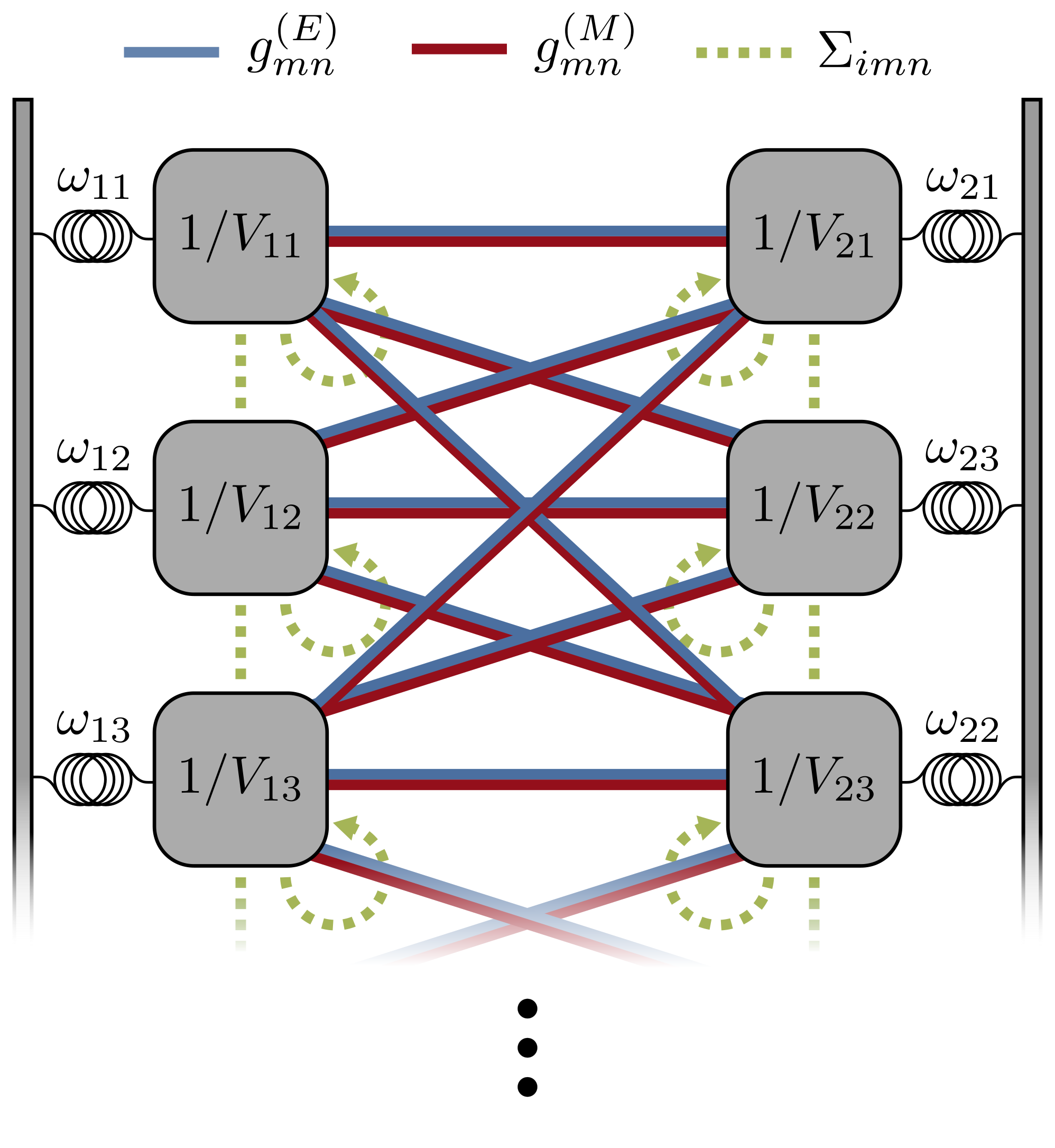}
\caption{A doubly-coupled oscillator model for electromagnetically interacting dielectric cavities. As shown in Eq.~\eqref{eq:Lagrangian}, the $m$th mode of the $i$th cavity is dynamically equivalent to a harmonic oscillator with natural frequency $\omega_{im}$ and mass $1/V_{im}$. When two or more cavities are brought into close proximity, three distinct types of couplings arise: the intracavity couplings $\Sigma_{imn}$, the intercavity magnetic couplings $g_{mn}^{(M)}$, and the intercavity electric couplings $g_{mn}^{(E)}$. Crucially, the simultaneous presence of the latter two suggests coupled cavity modes behave not as typical coordinate-coordinate coupled oscillators, but ``doubly'' coupled oscillators (i.e., involving independent couplings between both generalized coordinates and generalized velocities).}
\label{fig:f2}
\end{figure}

As illustrated in Figure \ref{fig:f2}, this Lagrangian is equivalent to that of a set of harmonic oscillators coupled via three distinct mechanisms: \\ \\
{(i) \emph{Intracavity coupling}}. Modes belonging to the same cavity are coupled through their electric fields according to the intracavity coupling strength $\Sigma_{imn}$. These terms appear due to the breakdown of orthogonality of the single cavity modes in the two cavity dielectric environment. As can be seen from Eq.~\eqref{eq:couplings}, the physical mechanism underlying this interaction of the single cavity electric field modes with the induced polarization of the dielectric media in regions where $\varepsilon(\mathbf{r})$ differs from the single cavity dielectric function $\varepsilon_i(\mathbf{r})$.\\ \\
{(ii) \emph{Intercavity electric coupling}}. Couplings terms scaling with $g_{mn}^{(E)}$ arise from the electric field portion of the Lagrangian between pairs of modes belonging to different cavities. Physically, they correspond to the interaction of the electric field of one cavity with the polarization induced by the field of the other. In the oscillator model, these terms manifest as interactions quadratic in the generalized velocities \\ \\
{(iii) \emph{Intercavity magnetic coupling}}. Couplings scaling with $g_{mn}^{(M)}$ arise from the magnetic field portion of the Lagrangian between pairs of modes belonging to different cavities. The form of $g_{mn}^{(M)}$ has been simplified using integration by parts ~\cite{smith2021theoretical}. In the oscillator model, these terms contribute couplings which are quadratic in the generalized coordinates. 

While it is well known that coupled cavity modes behave in analogy to coupled oscillators, the complex interplay of interaction terms which couple both the generalized coordinates and their time derivatives has not previously been appreciated to our knowledge. In typical applications of CMT, mode interactions are reduced to a form compatible with simple coordinate-coordinate coupling without inclusion of the additional contributions appearing in Eq.~\eqref{eq:Lagrangian}. Throughout the remainder of this manuscript, we will refer to the model derived here as doubly-coupled oscillators (DCOs), referring to the presence of both coordinate-coordinate and velocity-velocity coupling (or equivalently, as will be shown later, momentum-momentum coupling), in contrast with the ubiquitous model of coordinate-coupled oscillators (CCOs) analogous to a naive implementation of CMT. 

While we leave the physical consequences of this oversimplification to be discussed in the next section, it is important to remark that the preeminent work of Yariv et al. (Ref.~\cite{Yariv1999}), often cited to explain the underlying mechanism of coupling between adjacent cavities in works which utilize CMT, itself asserts the existence of three distinct coupling mechanisms. In the limit where the longitudinal corrections to the mode functions are ignored, it is straightforward to show that the three coupling parameters in Eq.~\eqref{eq:couplings} are identical to those derived in Ref.~\cite{Yariv1999} up to notational differences. In Sec.~\ref{sec:singlemode}, we will further elaborate on the impact these three couplings have on physical properties, such as those of the supermodes.

\section{Two single-mode cavities as doubly-coupled oscillators}\label{sec:singlemode}
Further analysis of the couple cavity Lagrangian is facilitated by simplification to the case of a photonic molecule dimer, with each cavity containing just a single mode within a spectral range of interest. This simplification is not strictly necessary, and much of the following discussion can be generalized for an arbitrary number of (spectrally isolated) cavity modes, but the interplay of the various interaction terms and its physical consequence is most digestible in this simplified form. 
In this limit, the coupled cavity Lagrangian becomes
\begin{equation}
\begin{split}
L &= \frac{1}{2}\sum_{i=1,2}\left[\frac{\dot{q}_i^2}{V_i}(1+\Sigma_i) - \frac{\omega_i^2}{V_i}q_i^2\right] \\
&+ \frac{g_{E}}{\sqrt{\omega_1\omega_2V_1V_2}}\dot{q}_1\dot{q}_2 - g_M\sqrt{\frac{\omega_1\omega_2}{V_1V_2}} q_1 q_2.
\end{split}
\label{eq:lagNotmatrixform}
\end{equation}
Similar to the more general case in Eq.~\eqref{eq:Lagrangian}, the single mode coupled cavity Lagrangian depends on the three distinct coupling parameters $\Sigma_i$, $g_E$ and $g_M$. Because only a single mode is considered in each cavity, the self-coupling scaling with $\Sigma_i$ may be compactly accounted for by replacing all quantities by their renormalized counterparts
\begin{equation}
\begin{split}
&\bar{V}_i = V_i/(1+\Sigma_i)\\
&\bar{\omega}_i = \omega_i/\sqrt{1+\Sigma_i} \\
&\bar{g}_E=g_E/[(1+\Sigma_1)(1+\Sigma_2)]^{3/4} \\
&\bar{g}_M=g_M/[(1+\Sigma_1)(1+\Sigma_2)]^{1/4}.
\end{split}
\end{equation}
Leveraging this notation, the Lagrangian may be written as
\begin{equation}
L=\frac{1}{2}\dot{\mathbf{q}}^T \mathbf{V}^{-1}\dot{\mathbf{q}} - \frac{1}{2} \mathbf{q}^{T}\mathbf{C}\mathbf{q},
\label{eq:lagMatrixform}
\end{equation}
where $\mathbf{q} = [\begin{matrix}q_1 & q_2 \end{matrix}]^T$ and
\begin{equation}
\begin{split}
\mathbf{V}^{-1}&=\left[\begin{matrix} 1/\bar{V}_1 & \bar{g}_E/\sqrt{\bar{\omega}_1\bar{\omega}_2\bar{V}_1\bar{V}_2} \\ \bar{g}_E/\sqrt{\bar{\omega}_1\bar{\omega}_2\bar{V}_1\bar{V}_2} & 1/\bar{V}_2\end{matrix}\right] \\
\mathbf{C} &=\left[\begin{matrix} \bar{\omega}_1^2/\bar{V}_1 & \bar{g}_M\sqrt{\bar{\omega}_1\bar{\omega}_2/\bar{V}_1\bar{V}_2} \\ \bar{g}_M\sqrt{\bar{\omega}_1\bar{\omega}_2/\bar{V}_1\bar{V}_2} & \bar{\omega}_2^2/\bar{V}_2\end{matrix}\right]. \\
\end{split}
\label{eq:barparams}
\end{equation}

For completeness, we also write the coupled cavity Hamiltonian (which will be used later in Sections~\ref{subsec:weakcoupling}-\ref{subsec:PUSC}), computed via Legendre transform of the coupled cavity Lagrangian:
\begin{equation}
    \begin{split}
    H &= \frac{1}{2}\sum_{i=1,2}\left[\widetilde{V}_i\, p_i^2 + \frac{\widetilde{\omega}_i^2}{\widetilde{V_i}}q_i^2\right]\\
    & -\widetilde{g}_E \sqrt{\frac{\widetilde{V}_1\widetilde{V}_2}{\widetilde{\omega}_1 \widetilde{\omega}_2}}p_1 p_2 + \widetilde{g}_M \sqrt{\frac{\widetilde{\omega}_1\widetilde{\omega}_2}{\widetilde{V}_1 \widetilde{V}_2}}q_1q_2 \\
    & = \frac{1}{2}\mathbf{p}^T \mathbf{V}\mathbf{p} + \frac{1}{2}\mathbf{q}^T \mathbf{C}\mathbf{q},
    \end{split}
    \label{eq:HDCO}
\end{equation}
where $\mathbf{p} = [\begin{matrix}p_1 & p_2 \end{matrix}]^T$. Importantly, the canonical momentum $p_i$ is \emph{not} equivalent to the mechanical momentum $p^{\textrm{mech}}_i = \dot{q}_i/\bar{V}_i$ due to the velocity-velocity coupling in $L$; instead, $\mathbf{p} = \mathbf{V}^{-1}\dot{\mathbf{q}}$ with $\mathbf{V}^{-1}$ a non-diagonal ``mass matrix''.
Furthermore, the renormalized parameters appearing in $H$ take the form 
\begin{equation}
\begin{split}
\widetilde{V}_i & = \bar{V}_i / (1-\bar{g}_E^2/\bar{\omega}_1\bar{\omega}_2) \\
\widetilde{\omega}_i &= \bar{\omega}_i/\sqrt{1-\bar{g}_E^2/\bar{\omega}_1\bar{\omega}_2} \\
\widetilde{g}_{E} &= \bar{g}_{E}/\sqrt{1-\bar{g}_E^2/\bar{\omega}_1\bar{\omega}_2} \\
\widetilde{g}_{M} &= \bar{g}_{M}/\sqrt{1-\bar{g}_E^2/\bar{\omega}_1\bar{\omega}_2},
\end{split}
\label{eq:tildparams}
\end{equation}
where the factors in the denominator mathematically arise from the determinant of $\mathbf{V}^{-1}$ and physically account for repeated interactions facilitated by the electric coupling $g_E$.

To illustrate the deviation between the above Lagrangian/Hamiltonian and those of the more typical CCOs, it is informative to analyze the equations of motion. Application of the Euler-Lagrange equations to $L$ (or the Heisenberg equations to $H$) yields
\begin{equation}
\mathbf{V}^{-1}\ddot{\mathbf{q}} = -\mathbf{C}\mathbf{q}.
\label{eq:matrixEOM}
\end{equation}
Of central importance here is the appearance of both a non-diagonal mass matrix $\mathbf{V}^{-1}$ (resulting from the coupling of the generalized velocities) and a non-diagonal coefficient matrix $\mathbf{C}$ (resulting from the coupling of the generalized coordinates), such that Eq.~\eqref{eq:matrixEOM} describes a pair of DCOs. Interestingly, a similar situation arises in the theory of interacting circuits which are coupled both capacitively and inductively \cite{vool2017introduction}.

To appreciate the distinction between DCOs and CCOs, it is helpful to repackage Eq.~\eqref{eq:matrixEOM} into a more intuitive form by left-multiplying by $\mathbf{V}$, resulting in the asymmetric coupled equations 
\begin{equation}
\frac{d^2}{dt^2}\left[\begin{matrix}q_1 \\ q_2\end{matrix}\right]=-\left[\begin{matrix}\Omega_1^2 & \mathcal{G}_{12} \\ \mathcal{G}_{21} &  \Omega_2^2\end{matrix}\right]\left[\begin{matrix}q_1 \\ q_2\end{matrix}\right],
\label{eq:effectiveEOM}
\end{equation}
where 
\begin{align}
    \Omega_i&=\sqrt{\widetilde{\omega}^2_i-\widetilde{g}_E\widetilde{g}_M}\label{eq:efffreq} \\ \mathcal{G}_{ij} &= \sqrt{\frac{\bar{\omega}_j\bar{V}_i}{\bar{\omega}_i\bar{V}_j}}(\widetilde{g}_M\widetilde{\omega}_i-\widetilde{g}_E\widetilde{\omega}_j)\label{eq:effcoupling}
\end{align} denote effective frequencies and (asymmetric) coupling coefficients, and tilded quantities are given by Eq. (\ref{eq:tildparams}).

The above \textit{effective} dynamical equations of motion now closely resemble coordinate-coupled oscillator equations. However, caution must be exercised in interpreting the physical system through this lens. For one, Eq.~\eqref{eq:effectiveEOM} shows that the effective resonant frequencies $\Omega_i$ are more complicated than their bare counterparts due to a complex interplay of all three coupling mechanisms. Furthermore, in addition to their complicated dependence on the basic quantities $\Sigma_{im}$, $g_E$, and $g_M$, the effective coupling coefficients $\mathcal{G}_{ij}$ themselves depend on the bare resonant frequencies $\omega_i$. Interestingly, there is also an asymmetry in the off-diagonal coupling coefficients when $\bar{\omega}_1\neq \bar{\omega}_2$, resulting either from a nonzero detuning between the bare frequencies, or non-negligible asymmetric rescaling from the self-couplings $\Sigma_1$ and $\Sigma_2$.\footnote{We note that there is also asymmetry resulting from inequality of $V_1$ and $V_2$; however, this asymmetry is also expected for CCOs, as can be confirmed by recognizing that $\mathcal{G}_{12}/\mathcal{G}_{21}\to V_1/V_2$ in the limit $g_E\to 0$.} Altogether, it is this complicated scrambling of bare frequencies and multiple coupling mechanisms which distinguishes DCOs from CCOs.

In the next section, we will show how the DCO model derived to this point can be leveraged to predict the properties of the supermodes. Furthermore, we will demonstrate that the distinction between DCOs and the more intuitive case of CCOs provides a first-principles understanding for observable effects on supermode properties, such as coupling-induced resonance frequency shifts \cite{Popovic2006}.

\subsection{Deriving supermode properties from first principles} \label{subsec:supermode}

As previously mentioned, one strategy to solve for the supermodes of the two single-mode cavity system under study is to solve the generalized Helmholtz equation
\begin{equation}
    \nabla\times\nabla\times\mathbf{f}_{\pm}(\mathbf{r}) = \frac{\omega_{\pm}^2}{c^2}\varepsilon(\mathbf{r})\mathbf{f}_{\pm}(\mathbf{r}),
    \label{eq:coupled_helmholtz}
\end{equation}
where $\varepsilon(\mathbf{r})$ is the dielectric function of the composite system, and the subscript $\pm$ denotes the two orthogonal supermodes, notation we adopt for the remainder of this paper. As before, the mode functions provide an expansion basis for the vector potential,
\begin{equation}
    \mathbf{A}(\mathbf{r},t) = \sum_{s=\pm}\frac{c\sqrt{4\pi}}{V_s}q_s(t)\mathbf{f}_s(\mathbf{r}),
\end{equation}
and the properties established in Eq.~\eqref{eq:modeproperties} consequently follow, with the composite dielectric function taking place of that of the single cavity. In principle, this strategy is both straightforward and exact. As previously discussed, however, solving Eq.~\eqref{eq:coupled_helmholtz} can be computationally expensive and, depending on the complexity of the system, completely prohibitive.

In this section, we demonstrate how solutions to Eq.~\eqref{eq:coupled_helmholtz} can be constructed from the single cavity mode functions. For clarity, we carry this out for the two single-mode cavity system currently under study, but emphasize that the procedure is generalizable to larger systems of more modes and cavities. Irrespective of the particular system, the basic recipe is as follows -- first diagonalize the effective equations of motion Eq.~\eqref{eq:effectiveEOM}. Next, infer from the diagonalizing transformation the corresponding mixture of individual cavity modes which form the supermodes. Once the supermodes have been determined, their properties follow. We now carry this procedure out for the single-mode cavity dimer of the previous section.

\subsubsection{Supermode resonant frequencies and the effective coupling strength} \label{subsubsec:supermode_diag}
To compute properties of the supermodes, we must first diagonalize Eq.~\eqref{eq:effectiveEOM}, here expressed compactly as 
\begin{equation}
    \ddot{\mathbf{q}} = -\mathbf{VC}\mathbf{q}.
\end{equation}
This is achieved through similarity transform with respect to $\mathbf{X} = \mathbf{TRS}$, where
\begin{equation}
\mathbf{T} = \left[\begin{matrix}(\mathcal{G}_{12}/\sqrt{\mathcal{G}_{12}\mathcal{G}_{21}})^{1/2} & 0 \\ 0 & (\mathcal{G}_{21}/\sqrt{\mathcal{G}_{12}\mathcal{G}_{21}})^{1/2}\end{matrix}\right]
\end{equation}
is a scaling (or squeezing) matrix which forces the couplings to be symmetric,
\begin{equation}
\mathbf{R} = \left[\begin{matrix}\cos\theta & -\sin\theta \\ \sin\theta &\cos\theta\end{matrix}\right]
\end{equation}
rotates the scaled coordinates into the supermode basis with mixing angle $\theta=\tan^{-1}(2\sqrt{\mathcal{G}_{12}\mathcal{G}_{21}}/(\Omega_1^2-\Omega_2^2))/2$ or, reexpressed in terms of renormalized bare parameters,
\begin{equation}
    \begin{split}
        \theta=\frac{1}{2}\tan^{-1}\left(\frac{2\sqrt{(\bar{g}_E^2+\bar{g}_M^2)\bar{\omega}_1\bar{\omega}_2-\bar{g}_E\bar{g}_M(\bar{\omega}_1^2+\bar{\omega}_2^2)}}{\bar{\omega}_1^2-\bar{\omega}_2^2}\right),
    \end{split}
    \label{eq:mixingangle_theta}
\end{equation}
and
\begin{equation}
\mathbf{S} = \left[\begin{matrix}\alpha_+ & 0 \\ 0 &\alpha_-\end{matrix}\right]
\label{eq:scalingmatrixS}
\end{equation}
encodes a final scaling transformation. While the choice of $\alpha_{\pm}$ has no effect on the transformed equations of motion, we will later find that a consistent definition of the mode volume based on the normalization condition Eq.~\eqref{eq:normalization} constrains us to a particular choice for $\alpha_{\pm}$. For the present discussion, we leave $\alpha_{\pm}$ unspecified, assuming only that $\mathbf{S}$ is positive-definite; these parameters will later be chosen such that the transformed mode functions are properly normalized. We note that both $\mathbf{S}$ and $\mathbf{T}$ are equivalent to single-mode squeezing transformations when expressed as a canonical transformation at the level of the Hamiltonian.

Using the composite transformation matrix $\mathbf{X}$, transforming the equations of motion into supermode coordinates via $\mathbf{q}_{\pm} = \mathbf{X}^{-1}\mathbf{q}$ then yields,
\begin{equation}
\frac{d^2}{dt^2}\left[\begin{matrix}q_+ \\ q_-\end{matrix}\right]=-\left[\begin{matrix}\omega_+^2 & 0 \\ 0 &  \omega_-^2\end{matrix}\right]\left[\begin{matrix}q_+ \\ q_-\end{matrix}\right],
\label{eq:diagonal_eom}
\end{equation}
where the supermode resonance frequencies are given by
\begin{equation}
\begin{split}
    \omega_{\pm}^2 &= \frac{\widetilde{\omega}_1^2 + \widetilde{\omega}_2^2}{2} - \widetilde{g}_E\widetilde{g}_M \pm \frac{1}{2}\sqrt{(\widetilde{\omega}_1^2 - \widetilde{\omega}_2^2)^2 + 4\mathcal{G}_{12}\mathcal{G}_{21}} \\
    &=\frac{\Omega_1^2 + \Omega_2^2}{2} \pm \frac{1}{2}\sqrt{(\Omega_1^2 - \Omega_2^2 )^2 + 4\mathcal{G}_{12}\mathcal{G}_{21}}.
    \label{eq:supermode_quadratic}
\end{split}
\end{equation}
From the above expressions, we can define the effective coupling strength,
\begin{equation}
    g_{\textrm{eff}} =\sqrt{\frac{\mathcal{G}_{12}\mathcal{G}_{21}}{\Omega_1\Omega_2}},
    \label{eq:geff}
\end{equation}
which characterizes the timescale of coherent energy exchange between the oscillators. To clarify this physical interpretation, it is helpful to note that in the limit $4\mathcal{G}_{12}\mathcal{G}_{21}/(\Omega_1 + \Omega_2)^4 \ll 1$ (here analogous to the rotating wave approximation -- see App.~\ref{app:simpsupermode}), the supermode frequencies are well-approximated by
\begin{equation}
    \omega_{\pm} \approx \frac{\Omega_1  + \Omega_2}{2} \pm \frac{1}{2}\sqrt{(\Omega_1 - \Omega_2)^2 + g_{\textrm{eff}}^2}.
\label{eq:supermode_linear}
\end{equation}
Thus, $g_{\textrm{eff}}$ characterizes the normal mode frequency splitting between supermodes with degenerate effective frequencies. We emphasize that $g_{\textrm{eff}}$ is functionally dependent on $g_E$, $g_M$, and $\Sigma_i$, distilling all three coupling mechanisms into a single parameter. Though dissipation is not included in the present discussion, it also serves as an appropriate comparison to the dominant rate of dissipation for determination of weak versus strong coupling \cite{Smith2020}. Furthermore, note that in the limit where $g_E\to 0$, we find $g_{\textrm{eff}}\to g_M$, thus reverting to the case of CCOs, as expected. A similar limit can be taken for the case of momentum-coupled oscillators ($g_{M}\to 0$). Crucially, the form in Eq.~\eqref{eq:geff} interpolates between these two cases and provides a singular measure of coupling strength for the more general case of DCO.

\begin{figure}
\centering
\includegraphics[width=\linewidth,]{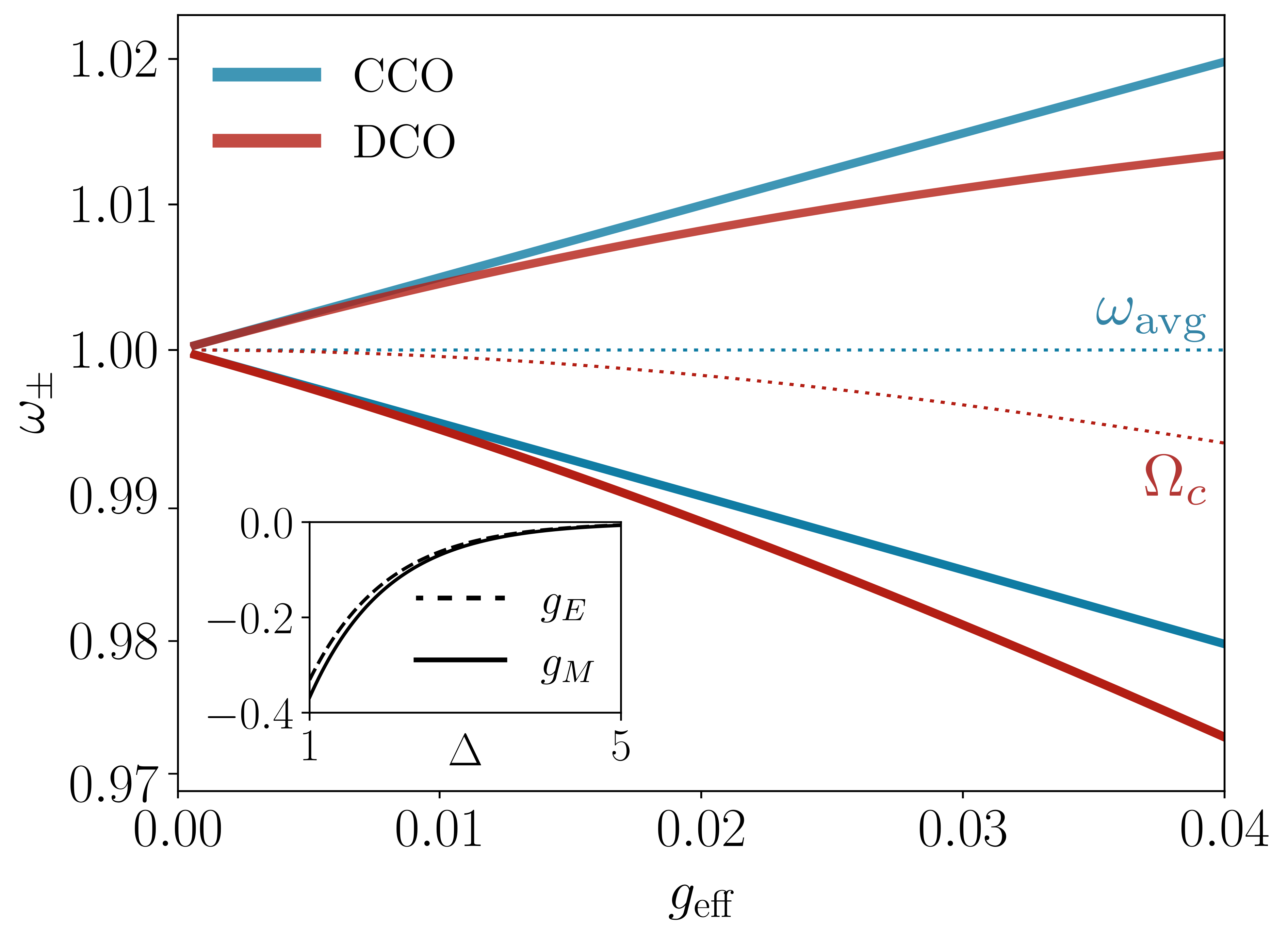}
\caption{Normal mode splitting in CCO and DCO models. For CCOs, the normal mode frequencies split about the average frequency $\omega_{\textrm{avg}} = (\omega_1 + \omega_2)/2$ (dashed blue line). In contrast, the normal modes of DCOs split about $\Omega_c$ (dashed red line), a complicated function of coupling parameters and bare frequencies that gives rise to coupling-induced frequency shifts -- see Eq.~\eqref{eq:BigOmegac}. To make these plots, we set $\Sigma_1=\Sigma_2 = 10^{-3}$ and $\omega_1 = \omega_2 = 1$. All frequencies and couplings are therefore in units of $\omega_{1,2}$. For the CCO model, we plot the conventional relation $\omega_\pm = \omega_{\textrm{avg}} \pm  g_{\textrm{eff}}/2$ for coordinate-coordinate coupling parameter $g_{\textrm{eff}}$ (see, e.g., Ref.~\cite{rodriguez2016classical}). The coupling parameters are varied according to $g_E = g_M/0.9 = -e^{-\Delta}$ for a fictitious control parameter $\Delta$, intended to model the dependence of cavity-cavity separation.}
\label{fig:f3}
\end{figure}

A few remarks are now in order regarding the properties of the supermodes for the case of DCO. As expected, we see that the supermode frequencies $\omega_+$ and $\omega_-$ are split about some central frequency $\Omega_{c}$. For the usual case of CCOs, $\Omega_c$ is the average of the bare resonant frequencies. For the coupled cavity mode system under study, however, we see that this is not the case. Instead, $\Omega_c$ is the average of the \emph{effective} frequencies,
\begin{equation}
    \Omega_c = \frac{\Omega_1+\Omega_2}{2} = \frac{1}{2}\left[\sqrt{\frac{\bar{\omega}_1^2 -\bar{g}_E\bar{g}_M}{1-\bar{g}_E^2/\bar{\omega}_1\bar{\omega}_2}}+\sqrt{\frac{\bar{\omega}_2^2 -\bar{g}_E\bar{g}_M}{1-\bar{g}_E^2/\bar{\omega}_1\bar{\omega}_2}}\right].
    \label{eq:BigOmegac}
\end{equation}
The above expressions, derived to full order in the intra- and inter-cavity couplings, provide a physical and analytical understanding of coupling-induced frequency shifts -- the phenomenon whereby normal mode frequencies split about a central frequency which itself depends on the coupling strength (see Fig. \ref{fig:f3}). In particular, we see that this effect arises not only from self interactions due to orthogonality breaking (scaling with $\Sigma_i$), as predicted in Ref.~\cite{Haus1991}, but also from electric ($g_E$) and magnetic ($g_M$) inter-cavity coupling terms. Momentarily specializing to the situation $\bar{\omega}_1=\bar{\omega}_2\equiv \bar{\omega}_0$, this is made especially clear by expanding $\Omega_c$ for small values of $\bar{g}_M/\bar{\omega}_0$ and $\bar{g}_E/\bar{\omega}_0$ up to second order, giving
\begin{equation}
    \Omega_c \approx \bar{\omega}_0 + \frac{1}{2}(\bar{g}_E/\bar{\omega}_0)(\bar{g}_E - \bar{g}_M),
    \label{eq:Omegac}
\end{equation}
Thus it is not only the contributions from $\Sigma_i$ which renormalize $\bar{\omega}_1$ and $\bar{\omega}_2$, but also the difference between electric and magnetic coupling terms which drives coupling-induced frequency shift phenomena. Perhaps more important than this quantitative understanding, though, is the intuition it provides: coupling-induced frequency shifts arise because coupled cavity modes behave not as simple CCOs, but as DCOs with interaction terms between generalized coordinates and their time derivatives.

The importance of the difference $\bar{g}_E - \bar{g}_M$ extends beyond its impact on the central frequency $\Omega_c$. The degree of splitting in the system is characterized by the generalized Rabi frequency $\Omega_{\textrm{Rabi}} = \sqrt{(\Omega_1-\Omega_2)^2 + g_{\textrm{eff}}^2}$ which itself is a complicated function of the bare system parameters. Again specializing to the case $\bar{\omega}_1=\bar{\omega}_2\equiv\bar{\omega}_0$, the Rabi frequency simplifies to the effective coupling strength which, in this limit, takes the simplified form
\begin{equation}
    \begin{split}
        g_{\textrm{eff}}   &=\frac{\widetilde{g}_M-\widetilde{g}_E}{\sqrt{1-\bar{g}_E\bar{g}_M/\bar{\omega}_0^2}} \approx \bar{g}_M-\bar{g}_E,
    \end{split}
    \label{eq:geff_v2}
\end{equation}
where, for the sake of intuition, we have expanded in small values of $\bar{g}_M/\bar{\omega}_0$ and $\bar{g}_E/\bar{\omega}_0$ to third order in the final approximation. The resulting expression provides a crucial insight into the supermode physics of photonic molecules -- the splitting in such systems is not determined by the electric or magnetic coupling alone, but rather their difference. In general, we find that $\bar{g}_E$ and $\bar{g}_M$ tend to have the same sign (see Section \ref{subsec:nanobeams} for one example, and Ref.~\cite{Smith2020} for another), thus allowing for the nonintuitive scenario where $\bar{g}_E$ and $\bar{g}_M$ are independently large in magnitude, but are low in contrast and therefore produce little splitting. In Section \ref{subsec:PUSC}, we demonstrate that such a situation gives elevated importance to counter-rotating terms in systems with relatively moderate splitting compared to the case of CCOs, thus suggesting the existence of interesting parameter regimes that lie beyond the traditional spectrum of strong, ultra-strong, and deep ultra-strong coupling.

\subsubsection{Supermode field profiles and mode volumes}\label{sssec:supermode}
We now demonstrate how this model can be used to analytically determine the supermode field profiles and related properties such as mode volumes, given knowledge of the single cavity field profile and the composite dielectric environment. For concreteness, we again specialize to the case of two single-mode cavities, but emphasize that the procedure is adaptable to many cavities, each with an arbitrary number of modes.

The key insight is to first notice that because the transformation matrix $\mathbf{X}$ diagonalizes the dynamical equations of motion (with spatial information integrated out), it must also diagonalize the wave equation (with spatial information intact), thereby providing a prescription to extract supermode field profiles. To see this, we first write the vector potential in the suggestive form
\begin{equation}
    \mathbf{A} = c\sqrt{4\pi}\,\mathbf{f}_V^T \mathbf{q}
    \label{eq:modeexp_vector}
\end{equation}
where $\mathbf{q} = [\begin{matrix}q_1 & q_2\end{matrix}]^T$ and $\mathbf{f}_V = [\begin{matrix}\widetilde{\mathbf{f}}_1(\mathbf{r})/V_1 & \widetilde{\mathbf{f}}_2(\mathbf{r})/V_2\end{matrix}]^T$, with the subscript $V$ indicating division by the mode volume. Inserting the identity $\mathbf{X}\mathbf{X}^{-1}$, the vector potential can be recast into the supermode basis:
\begin{equation}
    \begin{split}
        \mathbf{A}&=c\sqrt{4\pi}\,\mathbf{f}_V^T \mathbf{X}\mathbf{X}^{-1}\mathbf{q}\\
        &=c\sqrt{4\pi}\,\mathbf{f}_{V\pm}^T \mathbf{q}_\pm.
    \end{split}
    \label{eq:supermodeexp_vector}
\end{equation}
In other words, the fact that the coordinates transform according to $\mathbf{q}_{\pm}=\mathbf{X}^{-1}\mathbf{q}$ necessarily implies that the vector $\mathbf{f}_V$ transforms as $\mathbf{f}_{V\pm} = \mathbf{X}^T\mathbf{f}_V$, where $\mathbf{f}_{V\pm}=[\begin{matrix}\mathbf{f}_{+}(\mathbf{r})/V_+ & \mathbf{f}_-(\mathbf{r})/V_-\end{matrix}]^T$. In the case where $\mathbf{X}$ is unitary, $\mathbf{X}^T=\mathbf{X}^{-1}$ and both the coordinates and mode functions transform in an identical fashion. However, this need not be the case in general\footnote{We emphasize that nonunitarity of $\mathbf{X}$ does not imply equivalence with a noncanonical transformation. While we have chosen to diagonalize the system at the level of the classical equations of motion, the transformation matrix $\mathbf{X}$ can also be derived via purely unitary transformation of the corresponding (quantized) Hamiltonian. In general, the ability to express unitary transformation of a set of operators $\{O_i\}$ as $\hat{U}^\dagger \hat{O}_i \hat{U} = \sum_j X_{ij}\hat{O}_j$ does not imply $\mathbf{X}$ is unitary.} as the coefficient matrix $\mathbf{VC}$ is not guaranteed to be symmetric (e.g., when either $\bar{V}_1 \neq \bar{V}_2$ or $\bar{\omega}_1 \neq \bar{\omega}_2$).

To see that the transformed mode functions are indeed the solutions to the generalized Helmholtz equation in Eq.~\eqref{eq:coupled_helmholtz}, we express the generalized wave equation in terms of the supermode expansion Eq.~\eqref{eq:supermodeexp_vector}, yielding
\begin{equation}
    [\nabla\times\nabla\times \mathbf{f}_{V\pm}]^T\mathbf{q}_{\pm} + \frac{1}{c^2}\varepsilon(\mathbf{r})\mathbf{f}_{V\pm}^T\ddot{\mathbf{q}}_{\pm}=0.
\end{equation}
Leveraging the fact that the amplitudes $q_\pm$ are independent coefficients obeying $\ddot{q}_{\pm}=-\omega_{\pm}^2q_{\pm}$, it then follows that the mode functions $\mathbf{f}_{\pm}(\mathbf{r})$ are the sought-after solutions to Eq.~\eqref{eq:coupled_helmholtz} for the composite dielectric function $\varepsilon(\mathbf{r})$:
\begin{equation}
    \nabla\times\nabla\times \mathbf{f}_{\pm}(\mathbf{r}) = \frac{\omega_{\pm}^2}{c^2}\varepsilon(\mathbf{r})\mathbf{f}_{\pm}(\mathbf{r}).
    \label{eq:genHelm_super}
\end{equation}
Crucially, computation of these solutions requires only knowledge of the bare single cavity mode functions $\mathbf{f}_m(\mathbf{r})$ along with dielectric functions for the composite photonic molecule, $\varepsilon(\mathbf{r})$, and isolated single cavities, $\varepsilon_i(\mathbf{r})$, with all other quantities (e.g., $g_E$, $g_M$, etc.) being entirely determinable from this information.

In some sense, the system may now be viewed as a single dielectric cavity described by the composite dielectric function $\varepsilon(\mathbf{r})$. Correspondingly, we enforce all properties defined in Eq.~\eqref{eq:modeproperties} with $\varepsilon(\mathbf{r})$ replacing that of the single cavity. In particular, we note that the relation $\mathbf{f}_{V\pm} = \mathbf{X}^T\mathbf{f}_V$ only determines the supermode profiles $\mathbf{f}_{\pm}(\mathbf{r})$ up to an overall scaling (due to the fact that the diagonalizing transformation specifies the ratio $\mathbf{f}_{\pm}(\mathbf{r})/V_{\pm}$ rather than $\mathbf{f}_{\pm}(\mathbf{r})$ alone). For this reason, we have thus far left $\alpha_{\pm}$ (the elements of scaling transformation $\mathbf{S}$) undetermined. We now choose these coefficients to ensure $\textrm{max}\{\varepsilon(\mathbf{r})|\mathbf{f}_{\pm}|^2\}=1$ such that our desired normalization condition is met in analogy to Eq.~\eqref{eq:normalization}:
\begin{equation}
V_{\pm} = \frac{\int_{\mathcal{V}} d^3r\,\varepsilon(\mathbf{r})\left|\mathbf{E}_{\pm}(\mathbf{r})\right|^2}{\textrm{max} \{\varepsilon(\mathbf{r})\left|\mathbf{E}_{\pm}(\mathbf{r})\right|^2\}} = \int_{\mathcal{V}} d^3r\,\varepsilon(\mathbf{r})\left|\mathbf{f}_{\pm}(\mathbf{r})\right|^2,
\label{eq:supermodenormalization}
\end{equation}
where $\mathbf{E}_{\pm}(\mathbf{r})$ is the electric field contributed by the corresponding supermode. Carrying out this normalization, we find that the supermode functions are related to the modified mode functions $\widetilde{\mathbf{f}}_1(\mathbf{r})$ and $\widetilde{\mathbf{f}}_2(\mathbf{r})$ by 
\begin{widetext}
\begin{equation}
\begin{split}
\mathbf{f}_+(\mathbf{r}) &= \frac{1}{\mathcal{N}_+(\theta)}\left[  \left(\frac{\mathcal{G}_{12}}{\sqrt{\mathcal{G}_{12}\mathcal{G}_{21}}}\right)^{1/2}\sqrt{\frac{V_2}{V_1}}\,\widetilde{\mathbf{f}}_1(\mathbf{r})\cos\theta+ \left(\frac{\mathcal{G}_{21}}{\sqrt{\mathcal{G}_{12}\mathcal{G}_{21}}}\right)^{1/2}\sqrt{\frac{V_1}{V_2}}\,\widetilde{\mathbf{f}}_2(\mathbf{r})\sin\theta\right] \\ 
\mathbf{f}_-(\mathbf{r}) &= \frac{1}{\mathcal{N}_-(\theta)}\left[  \left(\frac{\mathcal{G}_{21}}{\sqrt{\mathcal{G}_{12}\mathcal{G}_{21}}}\right)^{1/2}\sqrt{\frac{V_1}{V_2}}\,\widetilde{\mathbf{f}}_2(\mathbf{r})\cos\theta-\left(\frac{\mathcal{G}_{12}}{\sqrt{\mathcal{G}_{12}\mathcal{G}_{21}}}\right)^{1/2}\sqrt{\frac{V_2}{V_1}}\,\widetilde{\mathbf{f}}_1(\mathbf{r})\sin\theta\right] \\
\end{split}
\label{eq:supermodefunctions}
\end{equation}
\end{widetext}
where the prefactors $\mathcal{N}_\pm(\theta)$ are inversely proportional to $\alpha_\pm$ and are defined in Appendix~\ref{subapp:explicit_forms}. 
Consistent with intuition, in the limit where $\theta \to 0$ (achieved by separating the cavities by a large distance such that $g_{\textrm{eff}}\to 0$, assuming $\widetilde{\omega}_1\neq\widetilde{\omega}_2$), we find that $\mathbf{f}_+(\mathbf{r})\to \widetilde{\mathbf{f}}_1(\mathbf{r})$ and $\mathbf{f}_-(\mathbf{r})\to \widetilde{\mathbf{f}}_2(\mathbf{r})$. This is in agreement with the expectation that the supermodes become equivalent to bare modes, and are thus each localized to a single cavity. In contrast, at maximal mixing $\theta\to \pi/4$, both $\mathbf{f}_{+}(\mathbf{r})$ and $\mathbf{f}_{-}(\mathbf{r})$ become superpositions of $\widetilde{\mathbf{f}}_1(\mathbf{r})$ and $\widetilde{\mathbf{f}}_2(\mathbf{r})$; therefore, the supermodes will generally be delocalized across the two cavities composing the photonic molecule. An exception to this occurs in cases when, for example, $V_1\ll V_2$ as the contribution of $\widetilde{\mathbf{f}}_1(\mathbf{r})$ to both $\mathbf{f}_+(\mathbf{r})$ and $\mathbf{f}_-(\mathbf{r})$ is proportional to $(V_2/V_1)^{1/4}$ when unraveled. In such cases, both mode functions are therefore localized to cavity 1 -- for an example of a system displaying this behavior, see Ref.~\cite{Smith2020}.

With the normalized supermode functions in hand, the supermode volumes $V_{\pm}$ can be computed directly via the integral relation in Eq.~\eqref{eq:supermodenormalization}, yielding

\begin{widetext}
\begin{equation}
\begin{split}
V_+ &= V_1\left[\frac{V_2}{V_1}\left(\frac{\mathcal{G}_{12}}{\sqrt{\mathcal{G}_{12}\mathcal{G}_{21}}}\right)\frac{1+\Sigma_1}{\mathcal{N}_+(\theta)^2}\right]\cos^2\theta+ V_2\left[\frac{V_1}{V_2}\left(\frac{\mathcal{G}_{21}}{\sqrt{\mathcal{G}_{12}\mathcal{G}_{21}}}\right)\frac{1+\Sigma_2}{\mathcal{N}_+(\theta)^2}\right]\sin^2\theta + \sqrt{V_1V_2}\left[\frac{g_E/\sqrt{\omega_1\omega_2}}{\mathcal{N}_+(\theta)^2}\right]\sin 2\theta \\
V_- &= V_2\left[\frac{V_1}{V_2}\left(\frac{\mathcal{G}_{21}}{\sqrt{\mathcal{G}_{12}\mathcal{G}_{21}}}\right)\frac{1+\Sigma_2}{\mathcal{N}_-(\theta)^2}\right]\cos^2\theta + V_1\left[\frac{V_2}{V_1}\left(\frac{\mathcal{G}_{12}}{\sqrt{\mathcal{G}_{12}\mathcal{G}_{21}}}\right)\frac{1+\Sigma_1}{\mathcal{N}_-(\theta)^2}\right]\sin^2\theta - \sqrt{V_1V_2}\left[\frac{g_E/\sqrt{\omega_1\omega_2}}{\mathcal{N}_-(\theta)^2}\right]\sin 2\theta. \\
\end{split}
\label{eq:supermodevolumes}
\end{equation}
\end{widetext}
Taking $V_{+}$ as an example, the first term derives from the integral $\int_{\mathcal{V}} d^3 r\,|\widetilde{\mathbf{f}}_1(\mathbf{r})|^2 = V_1(1+\Sigma_1)$, while the second incorporates a contribution from the second cavity, $\int_{\mathcal{V}} d^3 r\,|\widetilde{\mathbf{f}}_2(\mathbf{r})|^2 = V_2(1+\Sigma_2)$. Finally, the third term, proportional to $\int_{\mathcal{V}} d^3 r\, \widetilde{\mathbf{f}}_1(\mathbf{r})\cdot \widetilde{\mathbf{f}}_2(\mathbf{r}) = g_E\sqrt{\omega_1\omega_2/V_1V_2}$, accounts for interference between the two modes.

In the limit $\theta\to 0$, $V_+$ tends to $V_1(1+\Sigma_1)$ and not the bare mode volume $V_1$, as the former accounts for the modified dielectric background in the cavity dimer. Likewise, $V_-$ tends to $V_2(1+\Sigma_2)$. However, it is important to note that the limit $\theta\to 0$ is physically achieved by separating the two cavities by a large distance; in this case, $\Sigma_1\to 0$ and $\Sigma_2\to 0$, such that $V_\pm$ tend to the bare mode volumes, in agreement with expectation. 

To gain intuition for the opposite limit $\theta\to \pi/4$, it is helpful to consider the case of two identical cavities such that $V_1 = V_2 \equiv V_0$, $\omega_2 = \omega_1 \equiv \omega_0$, and $\Sigma_1 = \Sigma_2 \equiv \Sigma_0$. It can be shown that for the case where the two cavities are well-separated, the normalization factors can be approximated as $\mathcal{N}_{\pm}(\pi/4)\approx 1/2$ (see Appendix~\ref{subapp:explicit_forms}), leading to $V_{\pm}\approx 2V_0(1+\Sigma_0 \pm g_E/\omega_0)$. Thus, the supermode volumes are roughly double that of the bare modes up to (i) a correction scaling with the self-interaction $\Sigma_0$ and (ii) an interference term scaling with $g_E/\omega_0$. Importantly, this latter contribution is of opposite sign for the two normal modes: one experiences constructive interference, boosting the overall mode volume, while the other is characterized by destructive interference, reducing the mode volume. 

Between the two extreme limits $\theta=0$ and $\theta=\pi/4$, Eq.~\eqref{eq:supermodevolumes} captures a rich interplay of interference, coupling, and self-interaction effects. As alluded to above, a particularly interesting scenario arises for heterogeneous photonic molecules composed of cavities with drastically different mode volumes, as both supermodes can become localized to the same resonator. For more information, we refer to our prior work Ref.~\cite{Smith2020}.

As a final note, we emphasize that the analytic forms for the supermode functions and volumes provided in Eqs.~\eqref{eq:supermodefunctions} and ~\eqref{eq:supermodevolumes} are not only of fundamental interest, but are also practically useful. For example, they can be leveraged to make predictions about the coupling strength between the supermodes and a quantum emitter placed at a particular location without full electromagnetic simulations of the composite photonic molecule. By extension, this capability is useful for predicting observable effects such as Purcell enhancement~\cite{Purcell1946} -- dependent upon both the coupling strength and mode volume. This capability is particularly advantageous in systems where the coupling strength or other system parameters can be controlled (e.g., via optical~\cite{Sato2011}, mechanical~\cite{Siegle2016}, acousto-optic~\cite{Kapfinger2015}, electro-optic~\cite{Zhang2018}, or thermo-optic~\cite{Smith2020} methods), as one can analytically explore the realizable parameter space without the need for repeated simulations of the composite structure, opening up new pathways for lightweight and flexible design of photonic molecules for novel applications.

\subsection{The weak coupling limit: Reduction to coordinate-coupled oscillators} \label{subsec:weakcoupling}
In the previous section, we have shown that the physical description of a pair of coupled single mode cavities is equivalent to a doubly-coupled oscillator model. Furthermore, we have illustrated that the deviation of this DCO model from that of the ubiquitous CCO model gives rise to observable effects in important quantities such as the supermode frequencies. On the other hand, reduced-order modeling techniques such as CMT are often used to distill the physics of photonic molecules to either classical or quantum CCOs -- see, for example, Ref.~\cite{Liao_2020} for a review. It is well-known that such modeling techniques provide a valid description in the weak coupling limit~\cite{Haus1991}. Consequently, it stands to reason that the DCO model presented here must reduce to a CCO model in an appropriate weak-coupling limit. 

In this section, we show this to be the case. Motivated by this goal, we first consider a broader question: can the derived DCO model be unitarily transformed to an \emph{effective} CCO model? As both Hamiltonians are quadratic in coordinates and momenta, it is reasonable to expect this to be the case. Finding such a transformation not only provides intuition for the relationship between DCOs and CCOs, but the resulting effective CCO model answers a second pertinent question: if one naively fits experimental spectral data to a CCO model (e.g., using CMT), how are the fit parameters related to physical quantities? In other words, what is the corresponding physical Hamiltonian that is being fit? En route to showing that our DCO model Hamiltonian reduces to CCOs in the weak-coupling limit, we resolve these questions.

To begin, we recall the Hamiltonian for two single-mode cavities introduced in Eq. (\ref{eq:HDCO}),
\begin{equation}
    \begin{split}
    H_{\textrm{DCO}} & = \frac{\widetilde{V}_1}{2}p_1^2 + \frac{\widetilde{V}_2}{2}p_2^2 + \frac{\widetilde{\omega}_1^2}{2\widetilde{V}_1}q_1^2 + \frac{\widetilde{\omega}_2^2}{2\widetilde{V}_2}q_2^2 \\
    & -\widetilde{g}_E \sqrt{\frac{\widetilde{V}_1\widetilde{V}_2}{\widetilde{\omega}_1 \widetilde{\omega}_2}}p_1 p_2 + \widetilde{g}_M \sqrt{\frac{\widetilde{\omega}_1\widetilde{\omega}_2}{\widetilde{V}_1 \widetilde{V}_2}}q_1q_2,
    \end{split}
\end{equation}
here denoted with the subscript `DCO' to distinguish it from its `CCO' counterpart. To derive the latter from the former, we carry out a sequence of canonical transformations~\cite{Wagner1986,Merzbacher1998} characterized by the composite unitary operator $U = e^{S_1}e^{S_2}e^{S_3}$. The full procedure, along with the analytic form of the generators $S_1$, $S_2$, and $S_3$, is described in Appendix~\ref{app:effective Hamiltonian}. In brief, the first transformation (generated by $S_1$) diagonalizes $H_{\textrm{DCO}}$ while the second (generated by $S_2$) transforms from the diagonal Hamiltonian to one that includes only coordinate-coordinate coupling. When sequenced, these two non-commuting transformations enact a complicated mixture of beam-splitting, single- and two-mode squeezing as seen through the Baker-Hausdorff-Cambell formula. This suggests a complex relationship between DCOs and their effective CCO counterparts. Finally, the final transformation (generated by $S_3$) carries out a single-mode squeezing for each coordinate-coupled oscillator; the purpose of this final transformation is analogous to the role of the squeezing matrix $\mathbf{S}$ in Eq.~\eqref{eq:scalingmatrixS}, and its parameters are chosen such that the coordinates, mode functions, and effective mode volumes in the CCO frame are properly normalized (see the discussion surrounding Eq.~\eqref{eq:supermodenormalization} for related discussion, there for the supermode basis).

The result of this sequence of transformations is the first-principles, effective CCO Hamiltonian describing two single-mode coupled dielectric cavities\footnote{We note that in deriving this Hamiltonian, we have employed a passive (rather than active) transformation \cite{Merzbacher1998} such that $H_{\textrm{CCO}}$ is equivalent to $H_{\textrm{DCO}}$, but reexpressed in terms of the effective coordinates and momenta, $Q_i = U^\dagger q_i U$ and $P_i = U^\dagger p_i U$.}:
\begin{equation}
    \begin{split}
        H_{\textrm{CCO}} & = \frac{\mathcal{V}_{1}}{2}P_1^2 + \frac{\mathcal{V}_{2}}{2} P_2^2 + \frac{1}{2\mathcal{V}_{1}} \Omega_1^2 Q_1^2 +\frac{1}{2\mathcal{V}_{2}} \Omega_2^2 Q_2^2 \\ &+ g_{\textrm{eff}}\sqrt{\frac{\Omega_{1}\Omega_{2}}{\mathcal{V}_{1} \mathcal{V}_{2}}} Q_1Q_2.
    \end{split}
    \label{eq:effHamSCO}
\end{equation}
Here, $Q_i$ and $P_i$ are the effective coordinates and momenta in the CCO frame. 

We will analyze their analytic forms below for the special case of a homodimer (i.e., a system of two identical cavities); for the general case, see Appendix~\ref{app:effective Hamiltonian}. Furthermore, $\Omega_i$ and $\mathcal{V}_i$ are the effective frequency and mode volume for the $i$th mode; the former is defined in Eq.~\eqref{eq:efffreq}, while the latter is defined in the Appendix (see Eq.~\eqref{app:eq_eff_modevolume}). Notably, unlike the DCO model, here there is a single coupling term proportional to $g_{\textrm{eff}} = \sqrt{\frac{\mathcal{G}_{12}\mathcal{G}_{21}}{\Omega_1\Omega_2}}\approx \bar{g}_M - \bar{g}_E$, where the final approximation assumes the case of a homodimer. This is the \emph{effective} coupling strength first derived in Eq.~\eqref{eq:geff} and, as discussed, is related to the mode splitting -- see Eq.~\eqref{eq:geff_v2}. As expected, the effective CCO model naturally places this quantity at the forefront. Similarly, the effective frequencies $\Omega_i$ appearing in $H_{\textrm{CCO}}$ are identical to those derived via analysis of the equations of motion in Sec.~\ref{sec:singlemode}; see Eq.~\eqref{eq:effectiveEOM} in particular\footnote{It is interesting to note that one could have guessed the form of $H_{\textrm{CCO}}$ from the supermode frequencies in Eq.~\eqref{eq:supermode_quadratic}. Indeed, the algebraic manipulations carried out on the equations of motion in Sec.~\ref{subsubsec:supermode_diag} are  akin to the unitary transformations discussed in App.~\ref{app:effective Hamiltonian}.}.

From this result, it is tempting to conclude that while a DCO model naturally arises from first-principles, photonic molecules are just as well-described by the more ``typical'' case of CCOs.  However, this is not the case, as extreme caution must be exercised in interpreting $H_{\textrm{CCO}}$. As discussed in Sec.~\ref{sec:singlemode}, the effective frequencies $\Omega_i$ are complicated functions of coupling parameters ($g_E$, $g_M$ and $\Sigma_i$) and bare frequencies ($\omega_1$, $\omega_2$). In other words, the effective modes $Q_1$ and $Q_2$ do not correspond to the bare modes of the two cavities; instead, they are \emph{dressed} modes that incorporate complex effects induced by the altered two-cavity dielectric environment. This distinction is not only crucial for understanding effects beyond naive models such as coupling-induced frequency shifts \cite{Popovic2006} (see discussion around Eq.~\eqref{eq:Omegac}), but is necessary for interpreting system parameters estimated from experimental data. 

The nature of the dressed modes is further elucidated by inspecting the form of $Q_i$ and its corresponding mode function $\mathbf{F}_i(\mathbf{r})$. In particular, the effective coordinates are related to their bare counterparts via the dressing matrix ${\mathbf{M}}^{-1}$:
\begin{equation}
\left[\begin{matrix}Q_1 \\ Q_2\end{matrix}\right]=\mathbf{M}^{-1}\left[\begin{matrix}q_1 \\ q_2\end{matrix}\right],
\label{eq:dressing_relation}
\end{equation}
Likewise, the effective mode functions are related to those of the bare cavities by \begin{equation}
\left[\begin{matrix}\mathbf{F}_1(\mathbf{r})/\mathcal{V}_1 \\ \mathbf{F}_2(\mathbf{r})/\mathcal{V}_2\end{matrix}\right]=\mathbf{M}^{T}\left[\begin{matrix}\widetilde{\mathbf{f}}_1(\mathbf{r})/V_1 \\ \widetilde{\mathbf{f}}_2(\mathbf{r})/V_2\end{matrix}\right],
\label{eq:dressing_relation}
\end{equation}
in close analogy to the derivation of the supermode field profiles $\mathbf{f}_{\pm}(\mathbf{r})$ in Sec.~\ref{sssec:supermode}. As in Section~\ref{sssec:supermode} we narrow our focus on the simple scenario of a homodimer ($V_1 = V_2 \equiv V_0$, $\omega_2 = \omega_1 \equiv \omega_0$, and $\Sigma_1 = \Sigma_2 \equiv \Sigma_0$) and refer to Appendix~\ref{app:effective Hamiltonian} for the more general case. In this limit, the transformation matrix $\mathbf{M}$ takes the form
\begin{equation}
    \mathbf{M}^{-1} = 
    \begin{pmatrix} \alpha_1 & 0 \\ 0  & \alpha_2 \end{pmatrix}
    \begin{pmatrix} \cos\chi & \sin\chi \\ \sin\chi & \cos\chi \end{pmatrix},
\end{equation}
where the parameter $\chi$ is related to basic system parameters via
\begin{equation}
\chi = \lambda \tan^{-1}\left(\frac{1-\sqrt{1-(\bar{g}_E/\bar{\omega}_0)^2}}{\bar{g}_E/\bar{\omega}_0}\right),
\label{eq:chi}
\end{equation}
where $\lambda = (\bar{g}_M-\bar{g}_E)/|\bar{g}_M-\bar{g}_E|$ is an overall sign that arises due to our choice for positive square root sign convention, i.e., $\sqrt{x^2} = x$. Separately,  $\alpha_i$ are prefactors that scale the transformed modes such that the mode functions are properly normalized (see Appendix~\ref{app:effective Hamiltonian}), but otherwise do not impact the degree of hybridization between the bare cavity modes. We thus focus our discussion on right-hand matrix.

We first note that, interestingly, due to the matching sign on the off-diagonal terms, the right-hand matrix of $\mathbf{M}^{-1}$ is not a rotation matrix, but rather a non-orthogonal transformation matrix that is consistent with a pure two-mode squeezing~\cite{Wagner1986}.  Importantly, the parameter $\chi$ is independent of $g_M$, and is instead dependent upon the strength of the momentum-momentum coupling $g_E$ only. In the limit where $g_E\to0$, $H_{\textrm{DCO}}$ reverts to a purely coordinate-coupled Hamiltonian and, in agreement, we find $Q_i\to q_i$ and $\mathbf{F}_i(\mathbf{r})\to\widetilde{\mathbf{f}}_i(\mathbf{r})$. However, away from this limit, the effective modes described by $Q_i$ and $\mathbf{F}_i(\mathbf{r})$ are inequivalent to their bare counterparts and, instead, describe dressed, non-orthogonal modes that are \emph{delocalized} across the dimer. Furthermore, the modes become increasingly non-orthogonal with increasing $\bar{g}_E/\bar{\omega}_0 < 1$, illustrating the unsuitability of a ``naive'' CCO model to capture the essential physics for appreciable field overlap\footnote{We also note the pathological limit $\bar{g}_E/\bar{\omega}_0\to 1$ where the two modes coalesce, clearly demonstrating the important distinction between the bare cavity modes in the first-principles DCO model and the dressed modes of its effective CCO counterpart.}. Importantly, the effective CCO frame derived here is unique up to single-mode squeezings. In all, this suggests that one must be extremely careful in naively modeling strongly coupled photonic molecules with simple coordinate-coupled models, such as those commonly assumed in CMT. Indeed, we note that variants of CMT termed ``non-orthogonal CMT'' have been previously developed to capture such effects in strongly coupled resonators and waveguides ~\cite{Haus1991, Zhou2014}.

Finally, it is important to recognize that, while subtleties clearly arise for \emph{strongly} coupled photonic molecules, CCO models have been an essential and often successful tool for modeling weakly coupled photonic molecules~\cite{Liao_2020}. Thus, it stands to reason that in the appropriate limit, the DCO model should reduce to more ``typical'' coordinate-coupled oscillators. To see that this is indeed the case, we Taylor expand $\mathbf{M}^{-1}$ up to second order in $\bar{g}_E/\bar{\omega}_0$. Up to a normalization factor, this yields the following relationship between the effective and bare coordinates,
\begin{equation}
\begin{split}
Q_1 &= \left[1 - \frac{(\bar{g}_E/\bar{\omega}_0)^2}{8}\right]q_1 +\lambda \left[\frac{\bar{g}_E/\bar{\omega}_0}{2}\right]q_2 + O([\bar{g}_E/\bar{\omega}_0]^3) \\
Q_2 &= \left[1 - \frac{(\bar{g}_E/\bar{\omega}_0)^2}{8}\right]q_2 + \lambda \left[\frac{\bar{g}_E/\bar{\omega}_0}{2}\right]q_1 + O([\bar{g}_E/\bar{\omega}_0]^3)
\end{split}
\end{equation}
with an analogous relationship relating the mode functions $\mathbf{F}_i(\mathbf{r})$ and $\widetilde{\mathbf{f}}_i(\mathbf{r})$. Thus, for  $\bar{g}_E/\bar{\omega}_0 \ll 1$, the effective modes closely resemble those of the individual cavities, with only a weak dressing. If one discards this dressing, the subtleties of the effective frame dissolve and the system becomes identical to the more familiar case of coordinate-coupled oscillators.

\subsection{Emergence of pseudo-ultrastrong coupling from doubly-coupled oscillators}\label{subsec:PUSC}

In the previous section, we demonstrated that in an appropriately defined weak-coupling limit, the DCO model reduces to that of the more familiar CCO model commonly assumed in coupled mode theories. Here, we explore the opposite limit, revealing distinct behavior characterized by phenomena such as a squeezed vacuum ground state populated by virtual excitations. Such effects are typically associated with the ultrastrong coupling (USC) regime, where the coupling rate $g$ becomes a significant fraction of the system's natural frequencies ($g\gtrsim 0.1 \omega_0$), causing a breakdown of the rotating wave approximation ~\cite{Frisk_Kockum2019-pp, forn2019ultrastrong, qin2024quantum}. In this section, we show that the DCO model defies this classification due to the `decoupling' of co-rotating and counter-rotating terms in the Hamiltonian. This necessitates the definition of a regime we term \emph{pseudo-ultrastrong coupling} (pUSC), where the hallmark features of USC—such as a squeezed vacuum ground state with virtual excitations — emerge at comparatively modest mode splittings, offering new possibilities for experimental realization at optical frequencies.

Before continuing, we note that while USC has been closely studied for coupled linear oscillators \cite{peterson2019ultrastrong, PhysRevB.72.115303, PhysRevLett.121.040505, qin2024quantum}, much of the interest in USC physics over the past decade has been directed toward coupled light-matter systems comprising a single oscillator and a nonlinear element. Notable examples include a microwave cavity coupled to a transmon~\cite{bosman2017multi} or flux qubit~\cite{niemczyk2010circuit, forn2010observation}. While the presence of nonlinearity in these systems gives rise to additional non-classical effects beyond those captured by the purely linear model studied here, we emphasize that many of the hallmark phenomena of USC, such as virtual excitations in the vacuum state, are shared in common. Thus, while a full exploration is beyond the scope of our work, our findings remain relevant to the setting where one oscillator is replaced by a nonlinear element -- we give a few brief remarks on this possibility in Sec.~\ref{sec:conclusion}.

To begin, we quantize the two-mode Hamiltonian $H_\textrm{DCO}$ in Eq.~\eqref{eq:HDCO}. Invoking the canonical commutation relations $[q_i, p_j] = i \hbar \delta_{ij}$, we express generalized coordinates and momenta as
\begin{equation}
\begin{split}
q_{i} &= \sqrt{\hbar \widetilde{V}_i / 2\widetilde{\omega}_i} (a^\dagger_i + a_i) \\
p_{i} &= i \sqrt{ \hbar \widetilde{\omega}_i / 2\widetilde{V}_i}(a^\dagger_i - a_i),
\label{eq:qp_aadag}
\end{split}
\end{equation}
where $[a_i, a_{j}^\dagger] = \delta_{ij}$. Here, $a_i$ ($a_i^\dagger$) is the bosonic annihilation (creation) operator that lowers (raises) the photon number of the $i$th cavity mode, and the frequencies $\widetilde{\omega}_i$ and mode volumes $\widetilde{V}_i$ are the re-scaled parameters defined in Eq. (\ref{eq:tildparams}). We note that one can alternatively define the above relationship using bare parameters $\omega_i$, $V_i$ in place of the re-scaled counterparts $\widetilde{\omega}_i$, $\widetilde{V}_i$. Both conventions are related by a single-mode squeezing transformation and, importantly, choice of one over the other bears no impact on the eventual findings of this section. Thus, we opt for the definition in Eq.~\eqref{eq:qp_aadag} as it simplifies the mathematical expressions that follow. 

Casting $H_{\textrm{DCO}}$ in terms of $a_i$ and $a_i^{\dagger}$, we find
\begin{equation}
\begin{split}
    H_{\textrm{DCO}} & = \hbar \widetilde{\omega}_1 a^\dagger_1 a_1 + \hbar \widetilde{\omega}_2 a^\dagger_2 a_2 \\
    & + \hbar g_- (a^\dagger_1 a_2 + a_1 a^\dagger_2) + \hbar g_+(a_1 a_2 + a^\dagger_1 a^\dagger_2),
\end{split}
\label{eq:H_quantized}
\end{equation}
where $g_{\pm} = (\widetilde{g}_M \pm \widetilde{g}_E)/2$. It is helpful to contrast the above Hamiltonian with the more typical case of oscillators with coordinate-coordinate coupling. For the latter case, one finds
\begin{equation}
    \begin{split}
        H_{\textrm{CCO}} &= \hbar\omega_1 a_1^\dagger a_1 + \hbar\omega_2 a_2^\dagger a_2 \\
            & + \hbar g(a^\dagger_1 a_2 + a_1 a^\dagger_2 + a_1 a_2 + a^\dagger_1 a^\dagger_2),
    \end{split}
\end{equation}
Aside from some from relative minus signs in the interaction term, an identical form arises for Hamiltonians with a single momentum-momentum or momentum-coordinate coupling -- the latter naturally arising, for example, when modeling light-matter interactions in either the minimal coupling or dipolar gauge ~\cite{cohen1997photons}. Thus, aside from the rescaling of the frequencies $\widetilde{\omega}_i$, the primary distinction between doubly- and singly-coupled oscillators lies in the prefactors of the co-rotating ($a_1^{\dagger}a_2 + a_1 a_2^{\dagger}$) and counter-rotating terms ($a_1 a_2 + a_1^{\dagger}a_2^{\dagger}$) scale with different prefactors: in the singly-coupled case, there is one prefactor $g$ for both sets of terms, while in the doubly-coupled case the co-rotating and counter-rotating terms scale with \emph{distinct} parameters $g_+$ and $g_-$ that can take different values.

While both co-rotating and counter-rotating terms contribute to hybridization, the physical mechanism underlying each term is distinct. In particular, the effect of mode splitting can be traced back to the co-rotating terms. To see this in the DCO setting, note that the prefactor $g_- = (\widetilde{g}_M - \widetilde{g}_E)/2$ is closely related to the effective coupling strength $g_{\textrm{eff}}$, which itself is proportional to the vacuum Rabi frequency in the limit $\bar{\omega}_1 = \bar{\omega}_2= \bar{\omega}_0$. See Eq.~\eqref{eq:geff_v2} and the surrounding text. To make this connection concrete, in it can be shown that $g_- \approx \frac{1}{2}g_{\textrm{eff}}$ to third order in $\bar{g}_E/\bar{\omega}_0$ and $\bar{g}_M/\bar{\omega}_0$. 

Separately, the counter-rotating terms describe a two-mode squeezing interaction. For CCO systems, when the coupling strength $g$ is insignificant compared to the maximum resonance frequency, these terms can be discarded via the rotating wave approximation (RWA). In contrast, if $|g|/\textrm{max}\{\omega_1,\omega_2\} \gtrsim 0.1$, the RWA breaks down and the system is said to be ultrastrongly coupled. This manifests in a variety of interesting physical effects, most notably the presence of  entangled pairs of virtual photons in the vacuum. Crucially, $g$ is the prefactor for both co-rotating and counter-rotating terms. Realizing USC therefore requires one to engineer a system where the mode splitting is commensurate with the resonance frequencies -- a significant challenge attained thus far in only a few experimental platforms \cite{niemczyk2010circuit, bosman2017multi,PhysRevA.96.012325, PhysRevResearch.6.L042025,Baranov2020-cv,PhysRevLett.105.196402,Forn-Diaz2017-db}.

Contrasting with the DCO model, the co- and counter-rotating terms scale with distinct parameters $g_-$ and $g_+$, respectively. As a result, the definition of USC becomes murky -- the RWA breaks down when $g_+$ is commensurate with the resonance frequencies which, in principle, can occur independently of $g_-$. Thus, the normal mode splitting is decoupled from the ``turn-on'' of counter-rotating terms. It is this unique feature that motivates the definition of pUSC, which we define according to the condition 
\begin{equation}
    |g_+|/\textrm{max}\{\widetilde{\omega}_1,\widetilde{\omega}_2\} \gtrsim 0.1,
    \label{eq:pUSC}
\end{equation} 
consistent with the breakdown of the RWA\footnote{As a side remark, $|g_+|/\textrm{max}\{\bar{\omega}_1,\bar{\omega}_2\} \gtrsim 0.1$ is also a reasonable definition for pUSC. As these differ only at third order in $\bar{g}_E/\textrm{max}\{\bar{\omega}_1,\bar{\omega}_2\}$, we will use them interchangeably.}. 

Crucially, the pUSC regime of the DCO model captures the essential physics of the USC regime without the stringent requirement for extremely large coupling strengths. To demonstrate this, we now show that, similar to USC, pUSC is characterized by a ground state populated by virtual photons. To that end, we perform a sequence of unitary transformations to diagonalize Eq. (\ref{eq:H_quantized}), casting it in the form
\begin{equation}
    H_{\textrm{diag}} = \hbar \omega_+ a_+^\dagger a_+ + \hbar \omega_- a_-^\dagger a_-.
    \label{eq:H_quantized_diag}
\end{equation}
  See App.~\ref{app:VP} for details regarding the transformation. The supermode eigenfrequencies $\omega_{\pm}$ correspond to those previously derived in Eq.~(\ref{eq:supermode_quadratic}). Furthermore, the supermode annihilation operators $a_{\pm}$ can be expressed in terms of their bare counterparts via
\begin{equation}
    \begin{split}
        a_+ & = (\beta_1^{+}
        a_1+ \beta_1^{-} a_1^\dagger)\cos{\theta} + (\gamma_2^{+} a_2 + \gamma_2^{-} a_2^{\dagger})\sin{\theta} \\
        a_- & =  (\beta_2^{+}
        a_2+ \beta_2^{-} a_2^\dagger)\cos{\theta}-(\gamma_1^{+} a_1 + \gamma_1^{-} a_1^{\dagger})\sin{\theta},
    \end{split}
    \label{eq:a_pm}
\end{equation}
where $\theta$ is the mixing angle defined in Eq.~\eqref{eq:mixingangle_theta}. Noting that $a_+\to a_1$ and $a_- \to a_2$ in the limit $\theta\to 0$, we use the coefficients $\beta_i^{\pm}$ to denote ``diagonal'' contributions and $\gamma_i^{\pm}$ to indicate ``off-diagonal'' terms resulting from mode mixing. For the remainder of this section, we specialize to the simplified scenario of a homodimer ($V_1 = V_2 \equiv V_0$, $\omega_2 = \omega_1 \equiv \omega_0$, and $\Sigma_1 = \Sigma_2 \equiv \Sigma_0$); corresponding expressions for the more general setting of a heterodimer can be found in App.~\ref{app:effective Hamiltonian}. In this limit, the above coefficients take the form,
\begin{equation}
    \begin{split}
        \beta_1^{\pm} &= \frac{1}{2}\left(\zeta_{+} \pm \zeta_{+}^{-1}\right) \\ 
        \beta_2^{\pm} &= \frac{1}{2}\left(\zeta_{-} \pm \zeta_{-}^{-1}\right)  \\
        \gamma_{1}^{\pm} &= \lambda\beta^{\pm}_{2} \\
        \gamma_{2}^{\pm} &= \lambda\beta^{\pm}_{1},
    \end{split}
    \label{eq:a_pm_coeffs}
\end{equation}
where $\zeta_{\pm} = [(1\mp\bar{g}_M/\bar{\omega}_0)/(1\pm\bar{g}_E/\bar{\omega}_0)]^{1/4}$ and, similar to Eq.~\eqref{eq:chi}, $\lambda = (\bar{g}_M-\bar{g}_E)/|\bar{g}_M-\bar{g}_E|$ is an overall sign deriving from a choice in square root convention.

The virtual excitations in the vacuum are probed by computing the average occupancy of the bare modes in the supermode vacuum state \cite{Frisk_Kockum2019-pp, PhysRevB.72.115303}. For clarity, we denote the latter by $\ket{00}_{\pm}$ to avoid confusion with the ``false'' vacuum state of the bare cavities. Leveraging Eq.~\eqref{eq:a_pm}, a simple calculation then yields
\begin{equation}
    \begin{split}
        \langle 00 | a_i^{\dagger} a_i| 00\rangle_{\pm} &= [(\beta_i^-)^2 + (\gamma_i^-)^2]/2 \\
        &= \frac{1}{8}\left(\zeta_{+}^2 + \zeta_{-}^2 + 1/\zeta_{+}^2 + 1/\zeta_{-}^2\right) - \frac{1}{2}.
    \end{split}
    \label{eq:virtualpops}
\end{equation}
Here, the first line is general, while the second is particular to the case of a homodimer (see App.~\ref{app:VP}). Upon inspection, a few key features are immediately revealed. 

First, we see that the virtual photon population scales with contributions from $a_1^\dagger$ and $a_2^\dagger$ to the supermode operators $a_{\pm}$, each arising due to the non-negligible two-mode squeezing interaction in $H_{\textrm{DCO}}$. Because this interaction scales as $g_+ = (\widetilde{g}_M + \widetilde{g}_E)/2$, one would expect that the virtual populations disappear in the limit $g_E\to -g_M$ such that the RWA becomes exact. This is indeed the case, as $\beta^-_i\to 0$ and $\gamma^-_i\to 0$ in this limit\footnote{We note that this is true not only for the homodimer, but for more general setting of a heterodimer. See App.~\ref{app:VP} for details.}. 

Second, the DCO model can be reduced to the more familiar CCO model by taking the limit $\bar{g}_E \to 0$ (in this analogy, $\bar{g}_M\equiv g$ then becomes the sole coupling parameter). In turn, this simplifies $\zeta_{\pm}\to (1 \mp g/\bar{\omega}_0)^{1/4}$ such that the virtual photon population becomes
\begin{equation}
\langle 00 | a_i^{\dagger} a_i| 00\rangle_{\pm} = \frac{1}{16}[g/\bar{\omega}_0]^2 + O([g/\bar{\omega}_0]^4)\quad (\textrm{CCO})
\end{equation}
where we have Taylor expanded for small values of $g/\bar{\omega}_0$ to highlight the essential physics. Namely, we recover the well-established feature of CCOs that virtual occupancy of the ground states becomes meaningful only in the USC regime where $[g/\omega_0]^2$ becomes non-negligible.

Turning back to the more general DCO model, we find a parameter dependence beyond the conventional USC paradigm. To make the comparison clear, it is helpful to Taylor expand Eq.~\eqref{eq:virtualpops} about small values of $\bar{g}_E/\bar{\omega}_0$ and $\bar{g}_M/\bar{\omega}_0$. In turn, we find
\begin{equation}
\langle 00 | a_i^{\dagger} a_i| 00\rangle_{\pm} = \frac{1}{16}[g_+/\bar{\omega}_0]^2 + O([\bar{g}_x/\bar{\omega}_0]^4)\quad (\textrm{DCO}),
\end{equation}
where $O([\bar{g}_x/\bar{\omega}_0]^4)$ denotes a set of terms that are of total degree four in $\bar{g}_E/\bar{\omega}_0$ and $\bar{g}_M/\bar{\omega}_0$. Thus, for DCOs it is not the relative strength of the normal mode splitting (scaling with $g_- = (\widetilde{g}_M - \widetilde{g}_E)/2$) that is meaningful, but rather the independent coupling parameter $g_+=(\widetilde{g}_M + \widetilde{g}_E)/2$, motivating our definition of pUSC in Eq.~\eqref{eq:pUSC}. With this observation, we establish one of the primary results of this work: that phenomena conventionally associated with USC can be realized in DCO systems at comparatively moderate mode splittings, potentially opening the door for new experimental explorations. In the next Section, we provide an example of a simple system for which conventional USC is difficult to attain, yet pUSC is within reach for realistic experimental parameters.

\subsection{Example: Two coupled nanobeam resonators}\label{subsec:nanobeams}

\begin{figure*}
\centering
\includegraphics[scale=0.7]{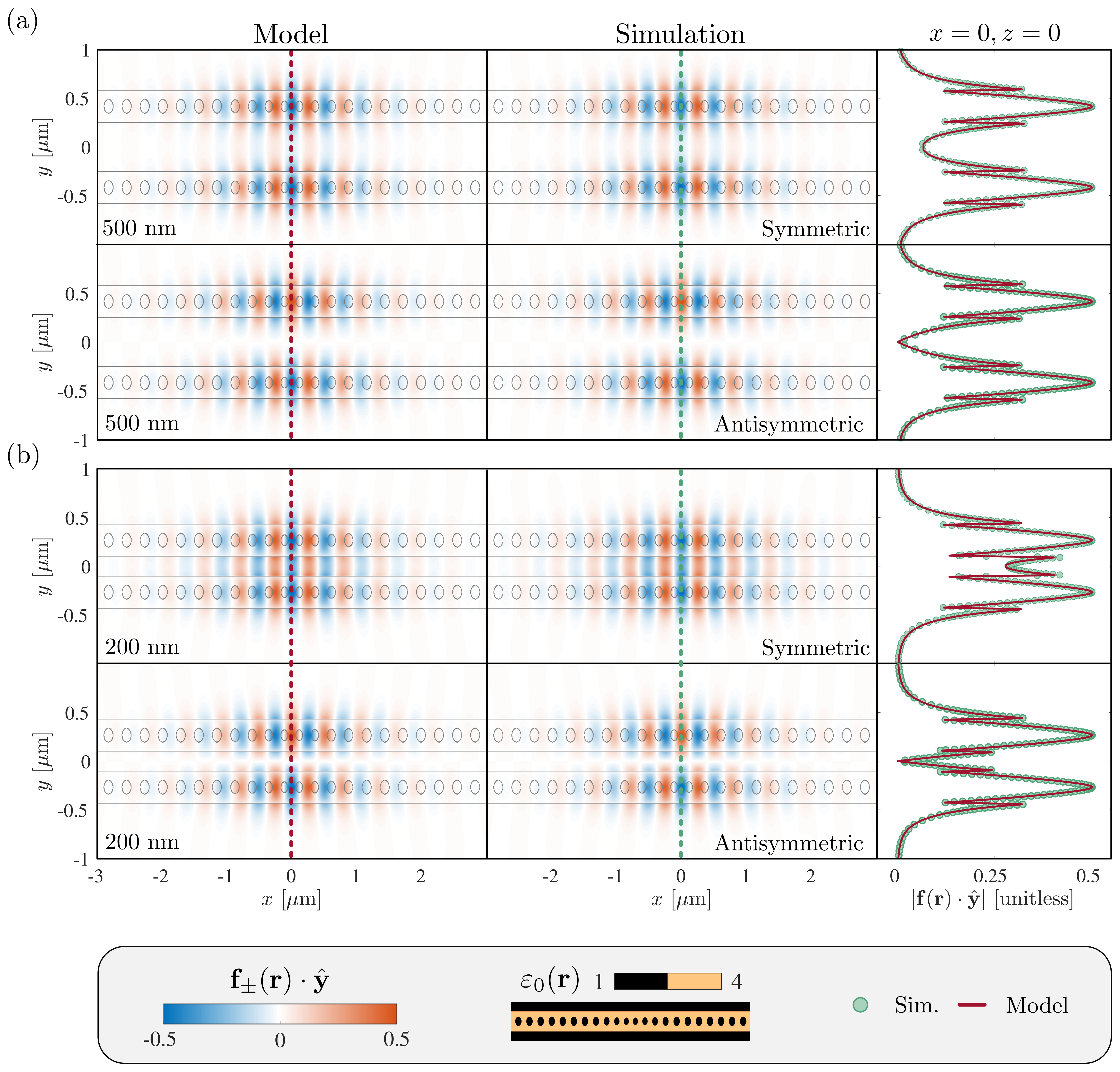}
\caption{The $\hat{\mathbf{y}}$-component of the supermode field profiles $\mathbf{f}_{\pm}(\mathbf{r})$ of a nanobeam-nanobeam homodimer. Panels (a) and (b) correspond to edge-to-edge cavity separations of 500 nm and 200 nm, respectively. Within each panel, the left plots show the model-predicted supermode, while the right display the result of FDTD simulations. Top rows correspond to the symmetric mode, $\mathbf{f}_-(\mathbf{r})$, with the bottom rows showing the the antisymmetric mode, $\mathbf{f}_-(\mathbf{r})$. For all plots, a two-dimensional cross-section at $z=0$ is displayed. Outsets at the right present one-dimensional cross-sections ($x=0$, $z=0$) comparing simulation and model, showing excellent agreement in all cases. For reference, the legend at the bottom includes the dielectric function of a single nanobeam cavity, $\varepsilon_0(\mathbf{r})$.}
\label{fig:f4}
\end{figure*}

We illustrate the practical utility of our theoretical framework by turning to a particular example: a homodimer composed of two silicon-nitride (SiN) photonic crystal nanobeam cavities ~\cite{deotare2009high, khan2011fabrication, fryett2018encapsulated}, each supporting a single mode within a large frequency window. In presenting this case-study, our aim is two-fold: First, given the properties of a single cavity, we demonstrate the ability to accurately predict supermode properties of the homodimer using the formalism presented in Secs. \ref{sec:lagrangian} and \ref{sec:singlemode}. Second, we show that the nanobeam-nanobeam homodimer is predicted to realize pUSC at relatively small mode splittings, opening up the possibility for experimental realization in a realistic platform.

To semi-analytically predict the properties of the nanobeam-nanboeam homodimer as a function of separation, we require the following inputs:
\begin{enumerate}
    \item The dielectric function for a single, isolated nanobeam resonator $\varepsilon_0(\mathbf{r})$.
    \item All relevant properties of the (single) nanobeam mode: the field profile $\mathbf{f}_0(\mathbf{r})$, natural frequency $\omega_0$, and mode volume $V_0$.
\end{enumerate}
We note that, strictly speaking, $\omega_0$, and $V_0$ can be inferred from $\varepsilon_0(\mathbf{r})$ and $\mathbf{f}_0(\mathbf{r})$ via the generalized Helmholtz equation Eq.~\eqref{eq:helmholtz} and normalization condition Eq.~\eqref{eq:normalization}, respectively. Here, we simplify matters by directly obtaining all properties $\mathbf{f}_0(\mathbf{r})$, $\omega_0$, and $V_0$ from finite-difference time-domain (FDTD) simulations. We use the dielectric function $\varepsilon_0(\mathbf{r})$ shown in the legend of Fig.~\ref{fig:f4}, corresponding to a single SiN nanobeam photonic crystal resonator with a 335 nm $\times$ 335 nm cross-section in vacuum. The field profile $\mathbf{f}_0(\mathbf{r})$ was computed on a computational domain terminated with perfectly matched layers (PMLs). For the single, high-$Q$ mode considered here, the near field in this computational region is an excellent approximation to the idealized PEC-enclosed domain $\mathcal{V}$.

To compute supermode properties of a nanobeam-nanobeam homodimer, we 
first compose two copies of $\varepsilon_0(\mathbf{r})$ to construct the composite dielectric function $\varepsilon(\mathbf{r})$ for a given cavity-cavity separation. Using this, we then solve for the modified, gauge-adjusted field profiles $\widetilde{\mathbf{f}}_i(\mathbf{r}) = \mathbf{f}_i(\mathbf{r}) + \nabla \psi_i(\mathbf{r})$, where $\mathbf{f}_i(\mathbf{r})$ is an appropriately shifted copy of $\mathbf{f}_0(\mathbf{r})$; en route, we compute the scalar corrections $\psi_i(\mathbf{r})$ by solving the generalized Poisson equation in Eq.~\eqref{eq:genpoisson} using fast-multipole methods \cite{fmmlib}. Finally, with the modified mode profiles $\widetilde{\mathbf{f}}_i(\mathbf{r})$ in-hand, we compute the three coupling parameters $\Sigma_0$, $g_E$, and $g_M$, each requiring the numerical evaluation of an integral; see Eq.~\eqref{eq:couplings}. Using the expressions derived in Sec.~\ref{subsec:supermode}, it is then straightforward to compute the supermode properties, such as the normal mode field profiles $\mathbf{f}_{\pm}(\mathbf{r})$, natural frequencies $\omega_{\pm}$, and mode volumes $V_{\pm}$. 

Fig.~\ref{fig:f4} shows a two-dimensional cross-section of the $\hat{\mathbf{y}}$-component of the symmetric and antisymmetric normal mode field profiles for separation distances of $200$ nm and $500$ nm. In particular, the left-hand column displays the field profile $\mathbf{f}_{\pm}(\mathbf{r}) \cdot  \hat{\mathbf{y}}$ at $z=0$ as predicted by our model via the semi-analytic procedure described above. For comparison, the right-hand column shows the result of simulations of the full composite structure. We find excellent agreement between the two, highlighting the power of our framework to predict complex properties like supermode field profiles by stitching together information from simpler simulations of individual components. While not shown here, the $\hat{\mathbf{x}}$-component shows similarly excellent agreement while the $\hat{\mathbf{z}}$-components are vanishingly small due to the geoemetric of the nanobeam resonator.

As a side note, we opt not to display the supermode volumes $V_{\pm}$ because they are uninteresting for homodimers; as noted below Eq.~\eqref{eq:supermodevolumes}, both $V_+$ and $V_-$ are roughly double the bare mode volume $V_0$ up to small corrections. For a scenario where the predicted supermode volumes display more interesting behavior, we refer to our earlier work on a ring-resonator-nanobeam heterodimer ~\cite{Smith2020}.

\begin{figure}
\centering
\includegraphics[width=\linewidth]{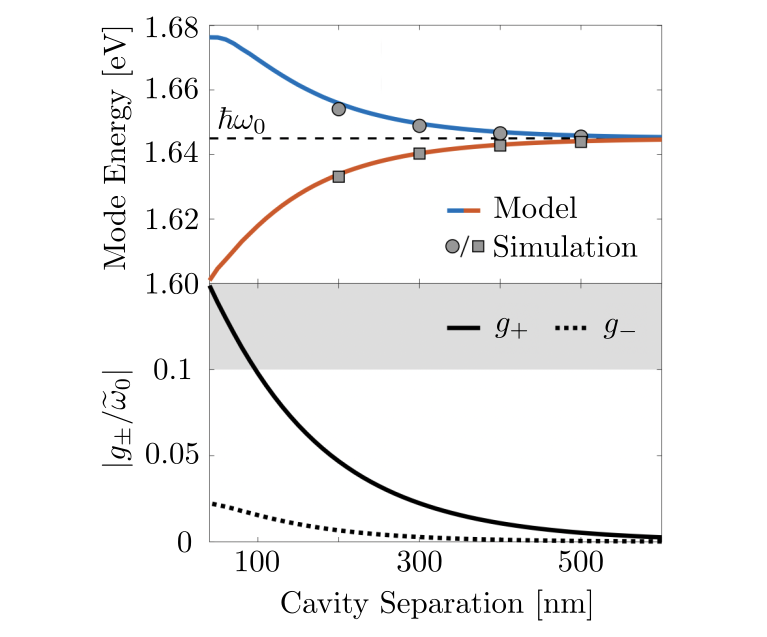}
\caption{Top panel: The supermode resonance energies $\hbar\omega_{\pm}$ as a function of cavity separation (edge-to-edge). Solid lines show $\hbar\omega_{\pm}$ as prediced by our model, while gray markers display the energies obtained via FDTD simulations of the full homodimer. The red (blue) line shows the resonance energy $\hbar\omega_-$ ($\hbar\omega_+$) for the symmetric (antisymmetric) mode. Bottom panel: model-extrapolated values of $g_\pm/\widetilde{\omega}_0$. The gray region indicated the onset of both USC ($|g_-/\widetilde{\omega}_0| \gtrsim 0.1$) and pUSC ($|g_+/\widetilde{\omega}_0| \gtrsim 0.1$), indicating the emergence of the latter for cavity separations $\lesssim$ 100 nm.}
\label{fig:f5}
\end{figure}

In the top panel of Fig.~\ref{fig:f5}, we compare the normal mode frequencies of the homodimer, reported in terms of energy per photon $\hbar\omega_{\pm}$. We clearly observe the influence of coupling-induced frequency shifts in both model and simulations, as $\omega_+$ and $\omega_-$ do not split symmetrically about the natural frequency $\omega_0$. The model again shows close agreement with FDTD simulations of the composite structure, though we note that we could not achieve convergence of the latter for separations below 200 nm.

The bottom panel of Fig.~\ref{fig:f5} shows the model-extrapolated values of $\eta_\pm = |g_\pm/\widetilde{\omega}_0|$ as a function of cavity separation. Here, the gray shaded region denotes the onset of pUSC (USC) for $\eta_+ \geq 0.1$ ($\eta_- \geq 0.1$). As foreshadowed in Sec. \ref{subsec:PUSC}, we observe that the coupled nanobeam cavities can be brought into pUSC at cavity separations of $\lesssim 100$ nm. Strikingly, for the same separation distance, the resonant coupling parameter $\eta_-$ is more than four times smaller than $\eta_+$. Quite notably, this disparity in magnitudes suggests the onset of USC-like effects for separations $\lesssim 100$ nm, despite relatively small normal mode splittings. Interestingly, we see that this separation between $\eta_-$ and  $\eta_+$ becomes even more pronounced at smaller cavity separations. 

These findings lead to a subtle yet crucial insight: a naive singly coupled oscillator model (e.g., CMT) would deem such mode splittings insufficient for USC-like phenomena. Put differently, fitting experimental data similar to the top panel of Fig.~\ref{fig:f5} using CMT would obscure the richness of the underlying physics, concealing the possibility of USC-like effects in pUSC. This interpretive gap underscores the limitations of singly coupled oscillator models in capturing the behavior of strongly coupled cavities and establishes the doubly coupled oscillator model presented here as a more powerful and rigorous framework for understanding and controlling the underlying physics. In addition, these findings establish pUSC as a simple alternative for realizing USC-type phenomena, such as virtual photons in a squeezed supermode vacuum, without the large mode splittings conventionally required for USC. This paves the way for experimental exploration of pUSC at optical frequencies and at room temperature in a practical dielectric resonator platform.

\section{Conclusion}\label{sec:conclusion}
In this work, we have developed a field-theoretic framework for modeling photonic molecules composed of two or more dielectric cavities within a finite perfect-electric-conductor (PEC) domain. This construction offers a controlled idealization of near-field cavity interactions and captures the behavior of spectrally isolated, high-$Q$ modes for which radiative corrections can be either neglected or incorporated perturbatively. Given only the properties of the individual cavities and their dielectric environment, this framework integrates Maxwell’s equations with Lagrangian mechanics to reduce the description of a strongly coupled photonic molecule to a set of effective coupled harmonic oscillators. Despite this apparent simplicity, our non-perturbative treatment of the cavity interactions reveals a complex interplay between intra- and inter-cavity coupling mechanisms that, in the effective coupled oscillator picture, give rise to emergent effects beyond those captured by CMT, which inherently relies on weak coupling assumptions. Thus, our framework serves as a powerful yet practical alternative to CMT, extending beyond its limitations while offering deeper physical insight.

A key practical advantage of our framework is its ability to predict supermode properties in coupled cavity systems without requiring full electromagnetic simulations of the composite structure. Such simulations, particularly FDTD methods, are computationally expensive and impractical for systematically exploring geometric dependencies like cavity-cavity separation and orientation, as each configuration requires a separate simulation. In contrast, our framework enables efficient assessment of coupling strengths and downstream near-field supermode properties, such as eigenfrequencies and mode profiles, using only simulations of the isolated components. We validate our PEC-based theoretical approach with a nanobeam resonator homodimer, demonstrating excellent agreement with full-structure simulations that use PMLs. While not explored here, we expect our framework to scale to large, heterogeneous multi-cavity photonic molecules, where traditional numerical electromagnetic simulations become increasingly prohibitive, provided each cavity contributes a small, spectrally isolated set of modes that are well approximated by PEC boundary conditions. By eliminating the need for costly full-system simulations, our framework provides a computationally efficient tool for designing and optimizing photonic molecules, with broad implications for applications ranging from quantum information processing to nonlinear optics. Looking ahead, additional techniques may be needed to generalize the present PEC-domain construction of coupled dielectric cavities to a full quasinormal-mode (QNM) formulation that rigorously combines both evanescent and propagating cavity modes through the inclusion of open asymptotic boundary conditions \cite{Kristensen2014, Kristensen2015, Sauvan_2022, wu2024exact}. However, if achieved, such a non-Hermitian extension would additionally enable predictions for far-field observables such as linewidths and hybridized $Q$-factors, while reducing to our results in the high-$Q$ limit in the absence of the continuous scattering spectrum.

Beyond its practical utility, our framework invites a refinement of the conventional understanding of coupled photonic modes. Specifically, we have shown that strongly coupled cavity modes are more accurately described as ``doubly coupled oscillators'' (DCOs) rather than the traditionally assumed coordinate-coupled oscillators (CCOs). This distinction is crucial: it not only explains the physical mechanism underlying previously observed phenomena without relying on ad hoc phenomenological parameters, but also clarifies the fundamental difference between weakly and strongly coupled photonic molecules. In the weak coupling regime where conventional CMT remains valid, the system behaves as a CCO, whereas in the strong coupling regime, the richer interaction structure of a DCO model becomes essential for capturing the full physics. In this work, we have made these distinctions rigorous by demonstrating that our model naturally reduces to a CCO description in the weak coupling limit ($\bar{g}_E/\bar{\omega}_0 \ll 1$), while deviations from this regime give rise to a DCO model with considerably distinct characteristics.

A defining consequence of this distinction is the emergence of a new parameter regime that we term pseudo-ultrastrong coupling. Like conventional ultrastrong coupling, pUSC is marked by a breakdown of the rotating wave approximation, giving rise to exotic phenomena such as virtual excitations in the supermode vacuum. However, unlike traditional ultrastrong coupling, pUSC does not require the experimentally demanding realization of extremely large mode splittings -- a direct consequence of the underlying DCO physics. This significantly lowers the barrier for accessing and utilizing phenomena typically associated with ultrastrong coupling, opening new avenues for experimental exploration in quantum optics, cavity QED, and beyond. Furthermore, we have presented a concrete example of a nanobeam homodimer, semi-analytically demonstrating that it can reach pUSC at relatively modest cavity-cavity separations, thereby paving the way for experimental investigation.

A natural question for future work is how to experimentally verify the existence of pUSC. To date, most works achieving USC have relied on the experimentally probed normal mode splittings to provide verification \cite{niemczyk2010circuit, bosman2017multi,PhysRevA.96.012325, PhysRevResearch.6.L042025,Baranov2020-cv,PhysRevLett.105.196402,Forn-Diaz2017-db} -- a strategy that, by definition, will not extend to pUSC. Instead, a promising alternative is the direct detection of virtual vacuum excitations, an approach that has garnered significant interest in conventional USC systems. Existing proposals involve the conversion of virtual photons into real photons via non-adiabatic modulation of system parameters~\cite{Frisk_Kockum2019-pp, PhysRevB.72.115303, minganti2024phonon, qin2024quantum}, stimulated Raman adiabatic passage to coherently amplify virtual excitations~\cite{falci2019ultrastrong}, and coherent control techniques that selectively extract virtual photons in superconducting circuits~\cite{giannelli2024detecting}. Exploring how these methods can be adapted to photonic molecules in the pUSC regime presents a promising path for experimental validation and a deeper understanding of strong light-matter interactions.

Another exciting direction for future exploration is the realization of DCO physics in alternative experimental platforms. One particularly promising candidate is circuit QED, where superconducting circuits can be engineered with simultaneous capacitive and inductive coupling -- a configuration that is both practical and highly controllable~\cite{blais2021circuit}. Interestingly, related ideas recently employed where dual charge and flux (i.e., coordinate and momentum) drives were used to exactly enforce the rotating wave approximation, enabling fast single-qubit gates~\cite{rower2024suppressing}. This represents, in some sense, a driven counterpart to the DCO framework presented here\footnote{We remark, however, that this idea is in some sense opposite to pUSC. Whereas pUSC relies on the enhancement of counter-rotating terms, Ref.~\cite{rower2024suppressing} effectively tunes the dual drive couplings to cancel the counter-rotating terms.}, further highlighting how dual coupling mechanisms -- whether between modes or to an external drive -- can be leveraged for precise control of interaction dynamics. Exploring how such engineered coupling schemes can be extended across quantum optics, superconducting circuits, and hybrid quantum systems presents an exciting avenue for future research.

\begin{acknowledgments}
We thank Arka Majumdar and Yueyang Chen for helpful discussions in the early stages of this work. We also thank Yueyang Chen for providing electromagnetic simulation data for the nanobeam-nanobeam homodimer. This research was supported by the National Science Foundation under award NSF CHE-1954393.
\end{acknowledgments}
\appendix
\setcounter{tocdepth}{1}
\section{The relationship between gauge, free charge, and the dielectric environment}
\label{app:gauge}
As discussed in the main text, there is an often understated connection between the choice of gauge and representation of both the dielectric environment and free charge. As these subtleties are essential to the both the consistency and novelty of our coupled cavity model, in this Appendix we explicitly show how two completely different macroscopic formulations of Maxwell's equations can be formed: one where all charge in the system is treated through a dielectric formalism and another where the charge is partitioned into ``dielectric'' and ``free'' charges, where only the former is packaged into a dielectric function. For clarity, all quantities which differ between the first and second formulations will be denoted by a superscript \RN{1} or \RN{2}, respectively. We will then show how these two formulations are related through a gauge transformation. This connection is not often made explicit in the literature, but has been hinted at in various works exploring quantization in dielectric media \cite{Dalton1996, Zietal2019}. The aim is for this section to provide an explicit, mathematical exposition to supplement the more qualitative discussion of Section \ref{subsec:gaugetrans} in the main text.
\subsection{Formulation \textrm{\Romannum{1}}}
In the presence of sources, Maxwell's equations are given by
\begin{equation}
\begin{split}
&\nabla\cdot\mathbf{E}=4\pi\rho \\
&\nabla\cdot\mathbf{B}=0 \\
&\nabla\times\mathbf{E} = -\frac{1}{c}\dot{\mathbf{B}} \\
&\nabla\times\mathbf{B}=\frac{4\pi}{c}\mathbf{j}+\frac{1}{c}\dot{\mathbf{E}}. \\
\end{split}
\label{eq:gauss}
\end{equation}
In the case where the charge density $\rho(\mathbf{r})$ and current density $j(\mathbf{r})$ are well-described by a macroscopic medium, it is beneficial to replace them with the polarization density $\mathbf{P}^{\RN{1}}(\mathbf{r})$ using the relations $\rho=-\nabla\cdot\mathbf{P}^{\RN{1}}$ and $\mathbf{j}=\dot{\mathbf{P}}^{\RN{1}}$, where contributions from the magnetization current density have been ignored as $\mu=1$ for all media of interest in this work. Assuming the medium to have a linear response $\mathbf{P}^{\RN{1}}(\mathbf{r})=\chi^{\RN{1}}(\mathbf{r})\mathbf{E}(\mathbf{r})$, Gauss's and Ampere's laws may be rewritten in the form
\begin{equation}
\begin{split}
&\nabla\cdot\varepsilon^{\RN{1}}\mathbf{E} = 0 \\
&\nabla\times\mathbf{B}=\frac{1}{c}\varepsilon^{\RN{1}}\dot{\mathbf{E}}
\label{eq:f1gauss}
\end{split}
\end{equation}
where the dielectric function is defined by $\varepsilon^{\RN{1}}(\mathbf{r})=1+4\pi\chi^{\RN{1}}(\mathbf{r})$. Because the entirety of the charge density $\rho$ was repackaged as polarizable media described by the total dielectric function $\varepsilon^{\RN{1}}$, specialization to the modified Coulomb gauge $\nabla\cdot\varepsilon^{\RN{1}}\mathbf{A}^{\RN{1}}=0$ ensures that only electromagnetic degrees of freedom remain. With this choice of gauge, Gauss's law becomes the generalized Laplace equation
\begin{equation}
\nabla\cdot\varepsilon^{\RN{1}}\nabla\phi^{\RN{1}}=0,
\label{eq:f1lap}
\end{equation}
$\phi^{\RN{1}}$ may be taken to be vanishing without loss of generality, and Ampere's law becomes the wave equation for the vector potential
\begin{equation}
\nabla\times\nabla\times\mathbf{A}^{\RN{1}}+\frac{\varepsilon^{\RN{1}}}{c^2}\ddot{\mathbf{A}}^{\RN{1}}=0.
\label{eq:f1wave}
\end{equation}
The remaining two Maxwell's equations are trivially satisfied through the definitions of the potentials.

\subsection{Formulation \Romannum{2}}
While the procedure of defining a dielectric function and arriving at Eqs. (\ref{eq:f1gauss}--\ref{eq:f1wave}) is typically presented unambiguously, a choice was nonetheless made in relating the macroscopic polarization density $\mathbf{P}^{\RN{1}}$ to the total charge density $\rho$. Imagine instead that we wish to partition the charge density into two distinct contributions, $\rho(\mathbf{r}) = \rho_d(\mathbf{r}) + \rho_f(\mathbf{r})$, where the atoms which contribute to $\rho_d$ are to be accounted for by the dielectric function $\varepsilon^{\RN{2}}(\mathbf{r})=1+4\pi\chi^{\RN{2}}(\mathbf{r})$, and $\rho_f$ is to be treated as ``free charge''. Likewise, the current density is split into two separate contributions as $\mathbf{j}(\mathbf{r}) = \mathbf{j}_d(\mathbf{r})+\mathbf{j}_f(\mathbf{r})$. Then, similar to the previous formulation, $\rho_d$ and $\mathbf{j}_d$ are related to the macroscopic polarization density through the relations $\rho_d = -\nabla\cdot\mathbf{P}^{\RN{2}}$ and $\mathbf{j}_d=\dot{\mathbf{P}}^{\RN{2}}$, where $\mathbf{P}^{\RN{2}}(\mathbf{r})=\chi^{\RN{2}}(\mathbf{r})\mathbf{E}(\mathbf{r})$.

Choosing to split up the charge and current densities in this way leads to the following form of Gauss's and Ampere's laws:
\begin{equation}
\begin{split}
&\nabla\cdot\varepsilon^{\RN{2}}\mathbf{E}=4\pi\rho_f \\
&\nabla\times\mathbf{B}=\frac{4\pi}{c}\mathbf{j}_f+\frac{1}{c}\varepsilon^{\RN{2}}\dot{\mathbf{E}}
\end{split}
\end{equation}
As before, the generalized Coulomb gauge ensures that the scalar potential $\phi$ depends only upon the free charge in the system. Under the present formulation, this choice is consistent with the condition $\nabla\cdot\varepsilon^{\RN{2}}\mathbf{A}=0$, which leads to the generalized Poisson equation
\begin{equation}
	\nabla\cdot\varepsilon^{\RN{2}}\nabla\phi=-4\pi\rho_f
\label{eq:f2poiss}
\end{equation}
and the sourced wave equation
\begin{equation}
\nabla\times\nabla\times\mathbf{A}^{\RN{2}}+\frac{\varepsilon^{\RN{2}}}{c^2}\ddot{\mathbf{A}}^{\RN{2}}=\frac{4\pi}{c}\mathbf{j}_f-\frac{\varepsilon^{\RN{2}}}{c}\nabla\dot{\phi}^{\RN{2}},
\label{eq:f2wave}
\end{equation}
which differ from Eqs.~(\ref{eq:f1lap} -- \ref{eq:f1wave}) by the inclusion of source terms contributed by the free charge and current. Furthermore, Eq.~\eqref{eq:f1lap} implies a nonzero scalar potential, resulting in an additional source term in Eq.~\eqref{eq:f1wave}.

\subsection{Connecting formulations through gauge transformation}
We will now show that the two parallel formulations presented above are related by a gauge transformation, and that the choice of representing charge either as ``free'' or belonging to some macroscopic dielectric function may be understood through this lens. While either formulation may be chosen as a starting point, we choose to begin in formulation \RN{1} and arrive at \RN{2} through transformation. Noting that $\phi^{\RN{1}}=0$, gauge transformation of the scalar and vector potential in formulation \RN{1} gives $\mathbf{A}' = \mathbf{A}^{\RN{1}} +\nabla\psi$ and $\phi' =  - \dot{\psi}/c$, where $\psi$ is an arbitrary to be determined scalar function. In order to satisfy the gauge condition of formulation \RN{2}, we require that $\nabla\cdot\varepsilon^{\RN{2}}\mathbf{A}'=0$, leading to the condition
\begin{equation}
	\nabla\cdot\varepsilon^{\RN{2}}\nabla\psi=\nabla\cdot(\varepsilon^{\RN{1}}-\varepsilon^{\RN{2}})\mathbf{A}^{\RN{1}}.
\label{eq:gaugecond}
\end{equation}
Taking a time derivative of both sides and using the relations $\phi' = -\dot{\psi}/c$ and $\mathbf{E}=-\dot{\mathbf{A}}^{\RN{1}}/c$, Eq.~\eqref{eq:gaugecond} becomes the generalized Poisson equation
\begin{equation}
\begin{split}
\nabla\cdot\varepsilon^{\RN{2}}\nabla\phi'&=\nabla\cdot(\varepsilon^{\RN{1}}-\varepsilon^{\RN{2}})\mathbf{E} \\
&=-4\pi\nabla\cdot(\mathbf{P}^{\RN{2}} - \mathbf{P}^{\RN{1}})\\
&=-4\pi\rho_f,
\end{split}
\label{eq:gpoiss}
\end{equation}
which is identical to the generalized Poisson equation obeyed by $\phi^{\RN{2}}$. Turning now to the vector potential, Eq.~\eqref{eq:f1wave} can be rewritten as
\begin{equation}
\nabla\times\nabla\times\mathbf{A}^{\RN{1}} + \frac{\varepsilon^{\RN{2}}}{c^2}\ddot{\mathbf{A}}^{\RN{1}}=(\varepsilon^{\RN{1}}-\varepsilon^{\RN{2}})\frac{\dot{\mathbf{E}}}{c},
\end{equation}
where the relation $\mathbf{E}=-\dot{\mathbf{A}}^{\RN{1}}/c$ has again been used. Rewriting this relation in terms of the gauge transformed vector potential $\mathbf{A}'$, we find
\begin{equation}
\begin{split}
\nabla\times\nabla\times\mathbf{A}' + \frac{\varepsilon^{\RN{2}}}{c^2}\ddot{\mathbf{A}}' &= \frac{4\pi}{c}(\dot{\mathbf{P}}^{\RN{1}}-\dot{\mathbf{P}}^{\RN{2}})-\frac{\varepsilon^{\RN{2}}}{c^2}\nabla\ddot{\psi} \\
&=\frac{4\pi}{c}\mathbf{j}_f-\frac{\varepsilon^{\RN{2}}}{c}\nabla\dot{\phi}'.
\end{split}
\label{eq:gwave}
\end{equation}

While it is not obvious, the generalized Poisson and wave equations together uniquely define $\phi'$ and $\mathbf{A'}$ \cite{Dalton1996}. Comparison of Eqs. \eqref{eq:gpoiss} and \eqref{eq:gwave} with Eqs. (\ref{eq:f2poiss} -- \ref{eq:f2wave}) then leads to the conclusion that $\phi'=\phi^{\RN{2}}$ and $\mathbf{A}'=\mathbf{A}^{\RN{2}}$, unambiguously proving that the formulations \RN{1} and \RN{2} are related via gauge transformation. More broadly, gauge transformations between Coulomb-like gauges of the form $\nabla\cdot \xi(\mathbf{r})\mathbf{A}=0$, where $\xi(\mathbf{r})$ is an arbitrary function, are intricately linked to the representation of free charge in a macroscopic dielectric formalism. At its most extreme limit, this procedure even allows for expressions derived the generalized Coulomb gauge ($\xi(\mathbf{r}) = \varepsilon^{\RN{1}}(\mathbf{r})$), in which quantization in the presence of a dielectric is simplified, to be mapped back to the true Coulomb gauge ($\xi(\mathbf{r}) = 1$) \cite{Zietal2019}.

\section{Additional details on supermode properties}
\label{app:supermode}
In this section, we provide additional details supporting the calculations of Section~\ref{sssec:supermode}. This includes the simplification of the supermode frequencies to arrive at Eq. (\ref{eq:supermode_linear}), and the explicit written forms for the normalization constants $\mathcal{N}_+(\theta)$ and $\mathcal{N}_-(\theta)$ appearing in Eqs. (\ref{eq:supermodefunctions}) and (\ref{eq:supermodevolumes}).

\subsection{Simplification of the supermode frequencies}
\label{app:simpsupermode}

First, we briefly illustrate how the approximate expressions for $\omega_{\pm}$ in Eq.~\eqref{eq:supermode_linear} may be obtained from the exact expressions for the squared supermode frequencies, $\omega_{\pm}^2$ in Eq.~\eqref{eq:supermode_quadratic}. There are two ways to go about such a simplification. The first involves recasting the second-order equations of motion (involving coordinates and momenta) into first-order equations of motion (involving either creation and annihilation operators or their classical equivalents \cite{cohen1997photons}). Naturally, the eigenvalues of the former  correspond to squared frequencies, while diagonalization of the latter yields the normal mode frequencies themselves. Typically, the rotating wave approximation -- where the coupling strength is assumed to be small relative to the bare resonant frequencies -- is applied. Diagonalization then leads to simple expressions for the normal mode frequencies, analogous in form to Eq.~\eqref{eq:supermode_linear} putting aside the complicated functional dependence of $\Omega_i$ and $g_{\textrm{eff}}$. Here, we demonstrate how the same approximation may be taken through algebraic arguments alone. 

Noting that $\Omega_i^2 = \sqrt{\widetilde{\omega}_i^2-\widetilde{g}_E\widetilde{g}_M}$, we start by first rewriting Eq.~\eqref{eq:supermode_quadratic} as
\begin{equation}
    \omega_{\pm}^2 = \frac{\Omega_1^2 + \Omega_2^2}{2} \pm \frac{1}{2}\sqrt{(\Omega_1^2-\Omega_2^2)^2 + 4\mathcal{G}_{12}\mathcal{G}_{21}}.
\end{equation}
Using the identity $2\Omega_1^2 + 2\Omega_2^2 = (\Omega_1 -\Omega_2)^2 + (\Omega_1 + \Omega_2)^2$ and rearranging the terms inside the square root, this can be reexpressed as
\begin{equation}
    \begin{split}
        \omega_\pm^2 = &\frac{(\Omega_1 + \Omega_2)^2}{4} + \frac{(\Omega_1 - \Omega_2)^2}{4} \\
        &\pm 2\left(\frac{\Omega_1 + \Omega_2}{2}\right)\sqrt{\frac{(\Omega_1-\Omega_2)^2}{4}+\frac{\mathcal{G}_{12}\mathcal{G}_{21}}{(\Omega_1+\Omega_2)^2}}\\
        =& \left[\frac{\Omega_1 + \Omega_2}{2}
        \pm \frac{1}{2}\sqrt{(\Omega_1-\Omega_2)^2+\frac{4\mathcal{G}_{12}\mathcal{G}_{21}}{(\Omega_1+\Omega_2)^2}}\right]^2\\
        &-\frac{\mathcal{G}_{12}\mathcal{G}_{21}}{(\Omega_1+\Omega_2)^2},
    \end{split}
\end{equation}
where we have completed the square in the second equality. In the limit where $\mathcal{G}_{12}\mathcal{G}_{21}/(\Omega_1+\Omega_2)^2 \ll (\Omega_1 + \Omega_2)^2/4$ (where the right-hand expression is the dominant contribution to $\omega_\pm^2$), the term outside of the square may be dropped, resulting in 
\begin{equation}
    \omega_\pm\approx \frac{\Omega_1 + \Omega_2}{2}
        \pm \frac{1}{2}\sqrt{(\Omega_1-\Omega_2)^2+\frac{4\mathcal{G}_{12}\mathcal{G}_{21}}{(\Omega_1+\Omega_2)^2}}.
\end{equation}
 This is identical to Eq.~\eqref{eq:supermode_linear} in the main text up to the additional approximation,
 \begin{equation}
    \begin{split}
        4\mathcal{G}_{12}\mathcal{G}_{21}/(\Omega_1+\Omega_2)^2 &\approx \mathcal{G}_{12}\mathcal{G}_{21}/\Omega_1\Omega_2,
    \end{split}
 \end{equation}
 also valid in the specified parameter regime, in addition to the assignment $g_{\textrm{eff}} = \sqrt{\mathcal{G}_{12}\mathcal{G}_{21}/\Omega_1\Omega_2}$. The approximations used to obtain this form are identical to the simplifications resulting from the rotating wave approximation.

\subsection{Explicit forms for \texorpdfstring{$\mathcal{N}_+(\theta)$}{Nplus} and \texorpdfstring{$\mathcal{N}_-(\theta)$}{Nminus}}\label{subapp:explicit_forms}

Eq.~\eqref{eq:supermodefunctions} makes explicit the relationship between the supermode functions $\mathbf{f}_{\pm}$ and the bare mode functions of each cavity $\mathbf{f}(\mathbf{r})$. Importantly, these supermode functions are normalized in accordance with Eq.~\eqref{eq:supermodenormalization} for the following choice of $\mathcal{N}_{\pm}(\theta)$: 
\begin{widetext}
 \begin{equation}
\begin{split}
\mathcal{N}_+(\theta)^2 &= \frac{1}{\alpha_+}\frac{V_1 V_2}{V_+^2}=\textrm{max} \left\{\varepsilon(\mathbf{r})\left[  \left(\frac{\mathcal{G}_{12}}{\sqrt{\mathcal{G}_{12}\mathcal{G}_{21}}}\right)^{1/2}\sqrt{\frac{V_2}{V_1}}\,\widetilde{\mathbf{f}}_1(\mathbf{r})\cos\theta+ \left(\frac{\mathcal{G}_{21}}{\sqrt{\mathcal{G}_{12}\mathcal{G}_{21}}}\right)^{1/2}\sqrt{\frac{V_1}{V_2}}\,\widetilde{\mathbf{f}}_2(\mathbf{r})\sin\theta\right]^2\right\}\\
\mathcal{N}_-(\theta)^2 &= \frac{1}{\alpha_-}\frac{V_1 V_2}{V_-^2}=\textrm{max} \left\{\varepsilon(\mathbf{r})\left[\left(\frac{\mathcal{G}_{21}}{\sqrt{\mathcal{G}_{12}\mathcal{G}_{21}}}\right)^{1/2}\sqrt{\frac{V_1}{V_2}}\,\widetilde{\mathbf{f}}_2(\mathbf{r})\cos\theta  -\left(\frac{\mathcal{G}_{12}}{\sqrt{\mathcal{G}_{12}\mathcal{G}_{21}}}\right)^{1/2}\sqrt{\frac{V_2}{V_1}}\,\widetilde{\mathbf{f}}_1(\mathbf{r})\sin\theta\right]^2\right\}.
\end{split}
\end{equation}
\end{widetext}
In the limit where the two cavities are identical (i.e., $V_1=V_2$, $\Sigma_1=\Sigma_2$, and $\omega_1=\omega_2$), the normalization factors simplify as
\begin{equation}
\mathcal{N}_\pm^2 = \frac{1}{2}\textrm{max}\{\varepsilon(\mathbf{r}) [\widetilde{\mathbf{f}}_1(\mathbf{r}) + \widetilde{\mathbf{f}}_2(\mathbf{r})]^2 \},
\end{equation}
where we have used the fact that $\theta = \pi/4$ for $\bar{\omega}_1 = \bar{\omega}_2$. For cavities that are well-separated, it is reasonable to expect that the maximum of the above function will be located at a point $\mathbf{r}$ located within one of the cavities and where the overlap $\widetilde{\mathbf{f}}_1(\mathbf{r})\cdot\widetilde{\mathbf{f}}_2(\mathbf{r})$ is small. Further approximating $\widetilde{\mathbf{f}}_i(\mathbf{r})\approx \mathbf{f}_i(\mathbf{r})$, this leads to $\mathcal{N}_{\pm}^2\approx 1/2$. This approximation was used to analyze the qualitative behavior of $V_{\pm}$ in the limit $\theta\to\pi/4$ below Eq.~\eqref{eq:supermodevolumes}.

\section{Transforming to the effective coordinate-coupled Hamiltonian}\label{app:effective Hamiltonian}
\begin{figure}
\centering
\includegraphics[width=0.65\linewidth]{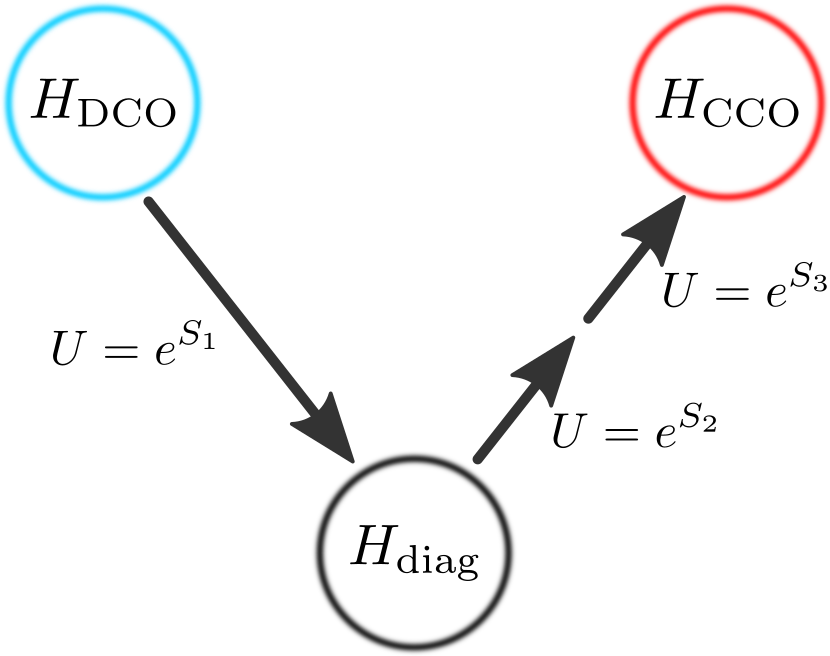}
\caption{Transformations between the doubly-coupled Hamiltonian $H_{\textrm{DCO}}$ and the effective coordinate-coupled Hamiltonian $H_{\textrm{CCO}}$. In total, three transformations are involved: (i) the first diagonalizes $H_{\textrm{DCO}}$, (ii) the second un-diagonalizes the Hamiltonian, introducing only a coordinate-coordinate coupling term, and (iii) the third performs a pair of single-mode  squeezing transformations that properly normalizes the modes and introduces ``proper'' mode volumes into the Hamiltonian. The coordinates and momenta corresponding to each of these frames are: (i) $( q_i, p_i ) \to ( q'_{\pm}, p'_{\pm} )$, (ii) $ ( q'_{\pm}, p'_{\pm} ) \to ( Q_i', P_i' )$, (iii) $( Q_i', P_i' ) \to ( Q_i, P_i )$. We note that we adopt primes on $( q'_{\pm}, p'_{\pm} )$ to distinguish from properly normalized supermodes, $( q_{\pm}, p_{\pm} )$ -- see discussion following Eq.~\eqref{eq:app_Vprime}}
\label{fig:f6}
\end{figure}
In this Appendix, we detail the diagonalization and subsequent transformation of the doubly-coupled Hamiltonian into an effective coordinate-coupled Hamiltonian. The aim is to construct a sequence of unitary transformations that, as shown in Fig.~\ref{fig:f6}, connects the two descriptions, thereby formalizing the equations of motion approach taken in Sec. \ref{subsubsec:supermode_diag}. We begin by briefly reviewing the method of unitary transformations, and refer to Ref.~\cite{Wagner1986} for more detail. The general strategy is to acquire a closed, analytic relationship between a canonical operator $\mathcal{O}$, and it's transformed counterpart $\widetilde{\mathcal{O}}$. This is accomplished via a unitary transformation operator $U = e^{S}$ and the Baker-Campbell-Hausdorff formula,
\begin{equation}
    \begin{split}
        \widetilde{\mathcal{O}} & = U^\dagger \mathcal{O} U \\
        & = \mathcal{O} + [\mathcal{O}, S] + \frac{1}{2!}[[\mathcal{O},S],S] +  \frac{1}{3!}[[[\mathcal{O},S],S],S] + ...,
    \end{split}
    \label{eq:app_BCH}
\end{equation} 
where the unitarity of $U$ preserves all commutation relations. Unitarity is satisfied if and only if the generating function $S$ is anti-Hermitian, $S^\dagger = -S$. 

Here, we consider successive transformations defined by generating functions of the common form,
\begin{equation}
    S = \frac{i \theta}{\hbar} \left(\alpha q_{j} p_{k} + \alpha' q_{j} p_{k}\right ).
\end{equation}
Intuitively, $S$ may be understood as simultaneously generating both beamsplitter and two-mode squeezing transformations, where the ratio of $\alpha$ and $\alpha'$ determine the strength of the latter and $\theta$ the former (assuming $\alpha = 1/\alpha'$, which will be the case for all transformations below). Not unexpectedly, we will see that the ``correct'' choice for $\theta$ will be the mixing angle defined in Eq.~\eqref{eq:mixingangle_theta}.

We begin with the double-coupled Hamiltonian in Eq.~\eqref{eq:HDCO}, restated here for convenience:
\begin{equation}
    \begin{split}
    H_{\textrm{DCO}} & = \frac{\widetilde{V}_1}{2}p_1^2 + \frac{\widetilde{V}_2}{2}p_2^2 + \frac{\widetilde{\omega}_1^2}{2\widetilde{V}_1}q_1^2 + \frac{\widetilde{\omega}_2^2}{2\widetilde{V}_2}q_2^2 \\
    & -\widetilde{g}_E \sqrt{\frac{\widetilde{V}_1\widetilde{V}_2}{\widetilde{\omega}_1 \widetilde{\omega}_2}}p_1 p_2 + \widetilde{g}_M \sqrt{\frac{\widetilde{\omega}_1\widetilde{\omega}_2}{\widetilde{V}_1 \widetilde{V}_2}}q_1q_2,
    \end{split}
    \label{eq:app_HamDCO}
\end{equation}
This Hamiltonian of Eq. (\ref{eq:app_HamDCO}) can be diagonalized using a unitary transformation generated by
\begin{equation}
    S_1 = \frac{i\theta}{\hbar}\left[\frac{\mathcal{G}_{21}}{\sqrt{\mathcal{G}_{12}\mathcal{G}_{21}}}q_1 p_2 - \frac{\mathcal{G}_{12}}{\sqrt{\mathcal{G}_{12}\mathcal{G}_{21}}}q_2 p_1  \right],
    \label{eq:app_SI}
\end{equation}
where $\theta$ is the mixing angle defined in Eq.~\eqref{eq:mixingangle_theta}. Defining $q_+' = e^{-S_1} q_1 e^{S_1}$, $q_-' = e^{-S_1} q_2 e^{S_1}$, and analogous relations for $p_\pm'$, this results in the Hamiltonian\footnote{We emphasize that while we adopt a subscript `diag' to differentiate between $H_{\textrm{DCO}}$ and the soon to be derived $H_{\textrm{CCO}}$ for clarity, we emphasize that all Hamiltonians are equivalent up to reexpression in terms of transformed operators. See Ref.~\cite{Wagner1986} for a useful discussion on the subtleties between active and passive transformations.}
\begin{equation}
    H_{\textrm{diag}} = \frac{\widetilde{V}_1^{'}}{2}p_{+}'^{2} + \frac{\widetilde{V}_{2}^{'}}{2}p_{-}'^{2} + \frac{1}{2\widetilde{V}_{1}^{'}}\omega_+^2 q_{+}'^{2} + \frac{1}{2\widetilde{V}_{2}^{'}}\omega_-^2 q_{-}'^{2},
    \label{eq:app_HamDCO_diag}
\end{equation}
Here, the (supermode) resonance frequencies $\omega_{\pm}$ are equivalent to those defined in Eq. (\ref{eq:supermode_quadratic}), while $\widetilde{V}_i'$ can be expressed in terms of basic system parameters as
\begin{equation}
    \begin{split}
        \widetilde{V}_{1}' & = \widetilde{V}_1\cos^2{\theta} + \widetilde{V}_2 \frac{ \mathcal{G}_{12}}{ \mathcal{G}_{21}} \sin^2{\theta} -\widetilde{g}_{E} \sqrt{\frac{\widetilde{V}_1 \widetilde{V}_2}{\widetilde{\omega}_1 \widetilde{\omega}_2  }}\frac{\mathcal{G}_{12}}{\sqrt{\mathcal{G}_{12} \mathcal{G}_{21}}}\sin{2\theta} \\
        \widetilde{V}_{2}' & =  \widetilde{V}_2\cos^2{\theta} + \widetilde{V}_1 \frac{ \mathcal{G}_{21}}{ \mathcal{G}_{12}} \sin^2{\theta} + \widetilde{g}_{E} \sqrt{\frac{\widetilde{V}_1 \widetilde{V}_2 }{\widetilde{\omega}_1 \widetilde{\omega}_2 }}\frac{\mathcal{G}_{21}}{\sqrt{\mathcal{G}_{12} \mathcal{G}_{21}} }\sin{2\theta}.
    \end{split}
    \label{eq:app_Vprime}    
\end{equation}
While tempting to interpret these parameters as the mode volumes of the supermodes, it is important to note that they are not the ``proper'' supermode volumes defined in Eq. (\ref{eq:supermodevolumes}). To see why, it is helpful to recall the equations of motion approach to diagonalization in Sec.~\ref{subsubsec:supermode_diag}. There, we utilized a final scaling transformation with parameters $\alpha_{\pm}$, chosen such the transformed mode functions are properly normalized according to Eq. (\ref{eq:supermodevolumes}). This is equivalent to performing a pair of single-mode squeezing transformations which ultimately rescale the mode volumes of the supermodes to their ``proper'' normalization. Here, we use ``proper'' in the sense that one should equivalently arrive at the same Hamiltonian and mode volumes if following the first-principles derivation of Sec.~\ref{ssec:single_cavity}, but treating the composite system as a single cavity with dielectric function $\varepsilon(\mathbf{r})$. For this reason, we have denoted the coordinate/momenta pairs by $(q_{\pm}',p_{\pm}')$ to distinguish them from their properly normalized counterparts $(q_{\pm},p_{\pm})$ appearing in Sec.~\ref{subsec:supermode}.

While we will transform to the ``proper'' supermode basis in Appendix~\ref{app:VP} (see Eq.~\eqref{eq:app_SSq} in particular), for the purpose of deriving an effective coordinate-coupled Hamiltonian the current frame is merely a pit-stop. Thus, we skip the normalization as it does not impact our final result, and follow the diagonalizing transformation $e^{S_1}$ by a second transformation generated by
\begin{equation}
    S_2 = -\frac{i\theta}{\hbar}\left[ \frac{\widetilde{V}_{2}'}{\sqrt{\widetilde{V}_{1}'\widetilde{V}_{2}'}} q_+' p_-' - \frac{\widetilde{V}_{1}'}{\sqrt{\widetilde{V}_{1}'\widetilde{V}_{2}'}}q_-' p_+'  \right].
    \label{eq:app_SII}
\end{equation}
Upon inspection of $S_1$ and $S_2$, it is evident that they are structurally similar with the only differences being (i) the squeezing parameters and (ii) the basis of the coordinates and conjugate momenta. Notably, the mixing angle $\theta$ is equivalent for both transformations. Expressed in terms of the transformed coordinates $Q_1'=e^{-S_2}q_+' e^{S_2}$, $Q_2'=e^{-S_2}q_-' e^{S_2}$ and their conjugate momenta, we find
\begin{equation}
    \begin{split}
        H_{\textrm{CCO}}' & = \frac{\widetilde{V}_{1}'}{2}P_1^2 + \frac{\widetilde{V}_{2}'}{2} P_2^2 \\
        & + \frac{1}{2\widetilde{V}_{1}'} \Omega_1^2 Q_1^2 +\frac{1}{2\widetilde{V}_{2}'} \Omega_2^2 Q_2^2 + g_{\textrm{eff}}\sqrt{\frac{\Omega_{1}\Omega_{2}}{\widetilde{V}_{1}' \widetilde{V}_{2}'}} Q_1Q_2,
    \end{split}
    \label{eq:app_effHamSCO}
\end{equation}
thus arriving at an effective Hamiltonian with solely coordinate-coordinate coupling. Note that the effective frequencies are equivalent to those derived via the equations of motion approach -- see Eq.~\eqref{eq:efffreq}. Likewise, the effective frequency $g_{\textrm{eff}}$ first defined in Eq.~(\ref{eq:geff}) and is a complicated repackaging of system parameters $g_E$, $g_M$, $\Sigma_i$, and $\omega_i$ that encodes the normal mode splitting. 

Similar to the discussion following Eq.~\eqref{eq:app_Vprime}, the parameters $\widetilde{V}_i'$ are not proper mode volumes corresponding to a normalized field profile. We therefore carry out a final single-mode squeezing transformation on each mode,
\begin{equation}
    \begin{split}
    S_3  = \frac{i}{2\hbar}\bigg[&\ln{\sqrt{|\widetilde{V}_{1}^{'}/\mathcal{V}_1}|} (Q_1' P_1' + P_1' Q_1') \\
    +&\ln{\sqrt{|\widetilde{V}_{2}^{'}/\mathcal{V}_2}|} (Q_2' P_2' + P_2' Q_2')\bigg],
    \end{split}
\label{eq:app_SIII}
\end{equation}
to arrive at properly normalized modes with effective coordinates $Q_i = U^{\dagger} q_i U$ and conjugate momenta $P_i = U^{\dagger} p_i U$, where $U=e^{S_1}e^{S_2}e^{S_3}$. Expressed in terms of these, the effective coordinate-coupled Hamiltonian then reads
\begin{equation}
    \begin{split}
        H_{\textrm{CCO}} & = \frac{\mathcal{V}_{1}}{2}P_1^2 + \frac{\mathcal{V}_{2}}{2} P_2^2 \\
        & + \frac{1}{2\mathcal{V}_{1}} \Omega_1^2 Q_1^2 +\frac{1}{2\mathcal{V}_{2}} \Omega_2^2 Q_2^2 + g_{\textrm{eff}}\sqrt{\frac{\Omega_{1}\Omega_{2}}{\mathcal{V}_{1} \mathcal{V}_{2}}} Q_1Q_2,
    \end{split}
    \label{eq:app_effHamSCO}
\end{equation}
where the effective mode volumes $\mathcal{V}_i$ are defined below.

In order to gain additional intuition into the relationship between the doubly- and coordinate-coupled Hamiltonians, one can analytically inspect the relationship between the effective and bare coordinates, $Q_i$ and $q_i$. We express this relation in the form $\mathbf{Q} = \mathbf{M}^{-1} \mathbf{q}$, where
$\mathbf{Q} = [\begin{matrix}Q_1 & Q_2 \end{matrix}]^T$ and $\mathbf{q} = [\begin{matrix}q_1 & q_2 \end{matrix}]^T$. The matrix $\mathbf{M}^{-1}$ connecting the original and effective pictures is
\begin{widetext}
\begin{equation}
    \mathbf{M}^{-1} = 
    \begin{pmatrix} \alpha_1 & 0 \\ 0  & \alpha_2 \end{pmatrix}
    \begin{pmatrix} \cos^2\theta + \mu_{21}\sin^2\theta & \frac{\mathcal{G}_{12}}{\sqrt{\mathcal{G}_{12}\mathcal{G}_{21}}}(1-\mu_{21})\sin\theta\cos\theta \\ -\frac{\mathcal{G}_{21}}{\sqrt{\mathcal{G}_{12}\mathcal{G}_{21}}}(1-\mu_{12})\sin\theta\cos\theta & \cos^2\theta + \mu_{12}\sin^2\theta \end{pmatrix},
    \label{eq:app_Mmatrix}
\end{equation}
\end{widetext}
where $\alpha_i = \sqrt{|\mathcal{V}_i/\widetilde{V}_i^{'}|}$ are the squeezing parameters that ensure proper normalization of the mode volumes, i.e., 
\begin{equation}
\mathcal{V}_i = \int_{\mathcal{V}} d^3 r\,
 \varepsilon(\mathbf{r})|\mathbf{F}_i(\mathbf{r})|^2,
 \label{app:eq_eff_modevolume}
 \end{equation}
 where the effective mode functions transform as $[\mathbf{F}_1(\mathbf{r})/\mathcal{V}_1 \quad \mathbf{F}_2/\mathcal{V}_2] =  [\widetilde{\mathbf{f}}_1(\mathbf{r})/V_1\quad  \widetilde{\mathbf{f}}_2/V_2]\,\mathbf{M}$, similar to Sec.~\ref{sssec:supermode}. Furthermore, we have adopted the shorthand,
\begin{equation}
\mu_{12} = \frac{\widetilde{V}_2^{'}}{\sqrt{\widetilde{V}_{1}^{'}\widetilde{V}_2^{'}}}\frac{\mathcal{G}_{12}}{\sqrt{\mathcal{G}_{12}\mathcal{G}_{21}}}, \quad \mu_{21} = \frac{\widetilde{V}_1^{'}}{\sqrt{\widetilde{V}_{1}^{'}\widetilde{V}_2^{'}}}\frac{\mathcal{G}_{21}}{\sqrt{\mathcal{G}_{12}\mathcal{G}_{21}}}.
\end{equation}
The effective momenta can be determined from the inverse transpose of this matrix which, importantly, preserves the symplectic form and, as a result, the canonical commutation relations between coordinates and momenta. This yields $\mathbf{P} = \mathbf{M}^{T} \mathbf{p}$, with $\mathbf{P} = [\begin{matrix}P_1 & P_1 \end{matrix}]^T$ and $\mathbf{p} = [\begin{matrix}p_1 & p_1 \end{matrix}]^T$. For a close analysis of $\mathbf{M}^{-1}$ in the homodimer limit, we refer to Sec.~\eqref{subsec:weakcoupling}.

\section{Supermode field operators and virtual photons in the supermode vacuum}\label{app:VP}
In this Appendix, we detail the calculation of the supermode creation and annihilation operators and the virtual photon population of the supermode vacuum. We begin with the diagonalized Hamiltonian in Eq. (\ref{eq:app_HamDCO_diag}), restated here for convenience:
\begin{equation}
    H_{\textrm{diag}} = \frac{\widetilde{V}_1^{'}}{2}p_{+}'^{2} + \frac{\widetilde{V}_{2}^{'}}{2}p_{-}'^{2} + \frac{1}{2\widetilde{V}_{1}^{'}}\omega_+^2 q_{+}'^{2} + \frac{1}{2\widetilde{V}_{2}^{'}}\omega_-^2 q_{-}'^{2}.
    \label{eq:app_HamDCO_diag1}
\end{equation}
By performing an additional pair of single-mode squeezing transformations generated by
\begin{equation}
\begin{split}
    S' = \frac{i}{2\hbar}\bigg[& \ln{\sqrt{\Big|\frac{\widetilde{V}_1'}{V_+}\Big|}} (q_+' p_+' + p_+' q_+') \\ 
    & + \ln{\sqrt{\Big|\frac{\widetilde{V}_2'}{V_-}\Big|}} (q_-' p_-' + p_-' q_-')\bigg],
\end{split}
\label{eq:app_SSq}
\end{equation}
we arrive at the diagonalized Hamiltonian,
\begin{equation}
    H = \frac{V_-}{2}p_{-}^{2} + \frac{V_+}{2}p_{+}^{2} + \frac{1}{2V_-}\omega_-^2 q_{-}^{2} + \frac{1}{2 V_+}\omega_+^2 q_{+}^{2},
    \label{eq:app_HDCORotSq}
\end{equation}
now expressed in terms of the ``proper'' supermode volumes defined in terms of the normalized field profiles via Eq. (\ref{eq:supermodevolumes}). We note that we have dropped the Hamiltonian subscript for simplicity of notation. The coordinates and conjugate momenta appearing in Eq. (\ref{eq:app_HDCORotSq}) are related to the supermode operators via 
\begin{equation}
    a_{\pm} = \sqrt{ \frac{\omega_{\pm}}{2\hbar V_{\pm}} }\left(q_{\pm} + \frac{iV_{\pm}}{ \omega_{\pm}}p_{\pm}\right).
\end{equation}
Further reexpressing the Hamiltonian in terms of these operators yields Eq.~\eqref{eq:H_quantized_diag} of the main text.

To probe the effects of the composite system ground state, it is useful to express the supermode operators in terms of their bare counterparts. Combining the relationships between the coordinates and momenta $(q_i, p_i)$ and the creation and annihilation operators ($a_{i}$, $a_{i}^\dagger$) given in Sec. (\ref{subsec:PUSC}) with the above, we arrive at the relationship,

\begin{equation}
    \begin{split}
        a_+ & = (\beta_1^{+}
        a_1+ \beta_1^{-} a_1^\dagger)\cos{\theta} + (\gamma_2^{+} a_2 + \gamma_2^{-} a_2^{\dagger})\sin{\theta} \\
        a_- & =  (\beta_2^{+}
        a_2+ \beta_2^{-} a_2^\dagger)\cos{\theta}-(\gamma_1^{+} a_1 + \gamma_1^{-} a_1^{\dagger})\sin{\theta}.
    \end{split}
    \label{eq:app_a_pm}
\end{equation}
Here, we have adopted notation that reflects the physical intuition behind each term. Each parameter $\beta_i^\pm$, $\gamma_i^\pm$ carries a subscript $i$ and superscript $+$ ($-$) that denotes contribution from the annihilation (creation) operator of the $i$th bare cavity mode. Noting that $a_+\to a_1$ and $a_- \to a_2$ in the limit $\theta\to 0$, we use $\beta_i^{\pm}$ to denote ``diagonal'' contributions and $\gamma_i^{\pm}$ to indicate ``off-diagonal'' terms resulting from mode mixing:
\begin{equation}
    \begin{split}
        \beta_1^\pm & = \frac{1}{2}\left(\sqrt{\frac{\omega_+ \widetilde{V}_1}{\widetilde{\omega}_1  \widetilde{V}_1'}} \pm \sqrt{\frac{\widetilde{\omega}_1  \widetilde{V}_1'}{\omega_+ \widetilde{V}_1}}\right) \\ 
        \beta_2^\pm & = \frac{1}{2}\left(\sqrt{\frac{\omega_- \widetilde{V}_2}{\widetilde{\omega}_2  \widetilde{V}_2'}} \pm \sqrt{\frac{\widetilde{\omega}_2  \widetilde{V}_2'}{\omega_- \widetilde{V}_2}}\right) \\
        \gamma_1^{\pm} &= \frac{1}{2\sqrt{\mathcal{G}_{12} \mathcal{G}_{21}}}\left( \mathcal{G}_{21}\sqrt{\frac{\omega_- \widetilde{V}_1}{\widetilde{\omega}_1  \widetilde{V}_2'}} \pm \mathcal{G}_{12} \sqrt{\frac{\widetilde{\omega}_1  \widetilde{V}_2'}{\omega_- \widetilde{V}_1}} \right) \\
        \gamma_2^{\pm} &= \frac{1}{2\sqrt{\mathcal{G}_{12} \mathcal{G}_{21}}}\left( \mathcal{G}_{12}\sqrt{\frac{\omega_+ \widetilde{V}_2}{\widetilde{\omega}_2  \widetilde{V}_1'}} \pm \mathcal{G}_{21} \sqrt{\frac{\widetilde{\omega}_2  \widetilde{V}_1'}{\omega_+ \widetilde{V}_2}} \right)
    \end{split}
    \label{eq:a_pm_coeffs}
\end{equation}
Likewise, the relations in Eq.~\eqref{eq:app_a_pm} can be inverted to express the bare mode operators in terms of their supermode counterparts:
\begin{equation}
    \begin{split}
        a_1 & = (\beta_1^{+}
        a_+ - \beta_1^{-} a_+^\dagger)\cos{\theta} - (\gamma_1^{+} a_- - \gamma_1^{-} a_-^{\dagger})\sin{\theta} \\
        a_2 & =  (\beta_2^{+}
        a_- - \beta_2^{-} a_-^\dagger)\cos{\theta} + (\gamma_2^{+} a_+ - \gamma_2^{-} a_+^{\dagger})\sin{\theta}.
    \end{split}
    \label{eq:app_a_pm}
\end{equation}

With the above analytic forms in hand, it is straightforward to compute the (virtual) photon population of the bare cavities in the vacuum state of the composite system, which we denote by $\ket{0_\pm}$. Proceeding with this calculation, we find
\begin{equation}
\bra{00_\pm} a_i^{\dagger} a_i \ket{00_\pm} = (\beta_i^-)^2\cos^2\theta + (\gamma_i^-)^2\sin^2\theta,
\end{equation}
in accordance with Eq.~\eqref{eq:virtualpops} of the main text, there simplified further for the case of a homodimer.

\bibliography{references}

\end{document}